\documentclass[useAMS,usenatbib]{mn2e}
\usepackage{amsfonts}
\usepackage{natbib}
\usepackage{float}
\usepackage{graphicx}                   
\usepackage{subfigure}

\title[PSR J0437$-$4715 profile stability]{Prospects for High-Precision Pulsar Timing}

\author[K.~Liu et al.]{K.~Liu,$^{1,2}$ J.~P.~W.~Verbiest,$^{2}$
  M.~Kramer,$^{1,2}$ B.~W.~Stappers,$^{1}$ W.~van Straten$^{3}$ \newauthor and J.~M.~Cordes$^{4}$\\
  $^{1}$University of Manchester, Jodrell Bank Centre for Astrophysics,
  Alan-Turing Building, Manchester M13 9PL, UK\\
  $^{2}$Max-Planck-Institut f\"{u}r Radioastronomie, Auf dem H\"{u}gel
  69, D-53121 Bonn, Germany \\
  $^{3}$Swinburne University of Technology, PO Box 218, Hawthorn VIC
  3122, Australia\\
  $^{4}$Astronomy Department, Cornell University, Ithaca, NY 14853,
  USA
  }

\begin{document}

\label{firstpage}

\maketitle

\begin{abstract}
  Timing pulses of pulsars has proved to be a most powerful technique
  useful to a host of research areas in astronomy and physics. Importantly,
  the precision of this timing is not only affected by radiometer noise, but
  also by intrinsic pulse shape changes, interstellar medium (ISM) evolution, instrumental distortions, etc.
  In this paper we review the known causes of pulse shape
  variations and assess their effect on the precision and accuracy of a single measurement of pulse arrival
  time with current instrumentation. Throughout this analysis we use the brightest and most
  precisely timed millisecond pulsar (MSP), PSR~J0437$-$4715, as a case study,
  and develop a set of diagnostic tools to evaluate profile stability in timing
  observations. We conclude that most causes of distortion can be either corrected
  by state-of-the-art techniques or taken into account in the estimation of time-of-arrival
  (TOA) uncertainties. The advent of a new generation of radio telescopes (e.g. the Square Kilometre Array, SKA),
  and their increase in collecting area has sparked speculation about the timing precision
  achievable through increases in gain. Based on our analysis of current data, we predict that for normal-brightness MSPs
  a TOA precision of between 80 and 230\,ns can be achieved at 1.4\,GHz with 10-minute integrations
  by the SKA. The actual rms timing residuals for each pulsar will be approximately at the same level only if
  all the other influences on timing precision (e.g. ISM, spin noise) are either corrected, modelled, or negligible.
\end{abstract}


\section{Introduction}
Pulsars are stable and rapidly rotating radio sources. This
stability \citep[which rivals the stability of atomic clocks on
Earth, see][]{hcmc10} can be exploited through pulsar timing. In
brief, pulsar timing works as follows: after the telescope surface
focusses the radio signal at the receiver feed, the signal is
amplified, sampled and digitised. Subsequently the
frequency-dependent dispersion delay caused by the ionised
interstellar medium (ISM) is removed in a process called
``de-dispersion''. Then a high signal-to-noise ratio (S/N) pulse
profile is obtained by averaging hundreds or thousands of subsequent
pulses in a step named ``folding''. Finally this high S/N average
profile is matched with an independently obtained standard profile
(or an analytic template based on the data itself), in order to
derive the time-of-arrival (TOA) of the integrated profile.

Precise monitoring of these TOAs allows the investigation of many
aspects of the pulsar, the binary system it inhabits and anything
affecting the radio wave propagation. For example, studies of strong
gravitational field effects in some binary pulsar systems have in
the past enabled some of the most constraining tests of general
relativity \citep{tw89,ksm+06}. Recent progress in software
\citep[e.g.][]{hjl+09} and hardware \citep[e.g.][]{vbb+10} has led
to predictions that within a decade, timing of a group of
millisecond pulsars (MSPs) will allow various characteristics of a
background of gravitational waves to be determined directly
\citep[e.g.][]{jhv+06,ljp08}. Furthermore, the next generation of
radio telescopes (such as the Square Kilometre Array, SKA, and the
Five hundred metre Aperture Spherical Telescope, FAST) will have the
best chance of detecting the first pulsar--black-hole binary system
\citep[e.g.][]{ckl+04}. Timing of such a system could determine the
spin and quadrupole moment of a black hole and then allow direct
tests of the Cosmic Censorship Conjecture and the no-hair theorem
\citep{kbc+04}. Pulsar timing at very high precision is required to
achieve the aforementioned scientific goals, and it is more readily
achieved with MSPs because of their short spin periods and highly
stable average pulse shapes. Currently, several MSPs have already
been timed at precisions down to a few hundred nanoseconds over time
spans of a decade or more \citep{vbc+09}.

Although the integrated profiles of MSPs appear stable over time
scales of years, there are a variety of effects that can affect the
shape of an integrated profile on short time scales: multi-path
propagation in the turbulent ISM, pulse jitter, data processing
artefacts and improper calibration, for example. Profile variations
from these effects may only change the pulse shape at low levels,
but will cause the subsequent TOA calculation to be less accurate
and precise than what is expected if only radiometer noise were
contributing to the uncertainty. This will complicate timing with
the next generation of radio telescopes since in these cases the
timing will be limited by factors other than merely telescope
sensitivity.

In order to investigate the level at which short-term instabilities
in pulse shape may affect pulsar timing with this new generation of
telescopes, we present an analysis on PSR~J0437$-$4715. This pulsar
was discovered by \cite{jlh+93} and is the nearest and brightest MSP
known, resulting in outstanding timing precision that has already
led to a variety of interesting results \citep{vbb+01,vbv+08}.
Furthermore, the TOA precision of PSR~J0437$-$4715 obtained by
current instruments \citep[see e.g.][]{vbb+10} is already comparable
to the precision future telescopes may expect to obtain on other
less bright MSPs (see Section \ref{sec:Conclusions}), making it a
perfect target for investigations of the pulsar timing potential of
future telescopes.

The structure of this paper is as follows. First we describe the
observations and data preprocessing in Section~\ref{sec:Obs}. Some
statistical tools are introduced in Section~\ref{sec:Tool}. Next we
review the possible effects involved in profile distortion and
present the results of data reduction in Section~\ref{sec:Issues}.
We conclude with an overview of our main findings, prospects for the
precision timing with the next generation of radio telescopes, and a
brief discussion of future research in
Section~\ref{sec:Conclusions}.

\section{Observations}\label{sec:Obs}
The data used in this paper consist of five long observations of
PSR~J0437$-$4715, taken between June 2005 and March 2008 at the
Parkes radio telescope. Observations were taken with the
Caltech-Parkes-Swinburne Recorder 2 \citep[CPSR2;][]{hbo06}. The
CPSR2 is a 2-bit baseband recorder that performs on-line coherent
dedispersion and records two 64-MHz wide observing bands
simultaneously. For the data used in this paper, these bands are
centred at observing frequencies of 1341 and 1405\,MHz. It also
effectively removes RFI online by monitoring the total power on
$\mu$s timescales and does not record any data whenever the power
levels deviate significantly from a Gaussian. On two of the five
days the data were taken with the H-OH receiver, on the remaining
three days the central beam of the 20\,cm multibeam (MB) receiver
\citep{swb+96} was used, as listed in Table~\ref{tab:Data}. During
each day of observations the data were folded in near-real time to
16.8\,s for the early data and to 67.1\,s for the later data (see
Table \ref{tab:Data}). Off-source observations of a pulsed noise
probe at 45$^{\circ}$ to the linear feed probes but with otherwise
identical set-up, were taken at regular intervals to allow for
polarimetric calibration.

For the data processing we used the \textsc{PSRchive} software
package \citep{hvm04}. We removed 12.5\,\% of each edge of the
bandpass to avoid possible effects of aliasing and spectral leakage.
Two models named ``single axis'' and ``full reception'',
respectively, were used for calibration purposes and details will be
presented in Section~\ref{ssec:Calib}. Unless otherwise specificied
in the text, we combined the polarisations into total power (Stokes
I) and the power across the remaining 96 frequency channels. Through
the following analysis, TOAs and their uncertainties were determined
through the standard cross-correlation approach \citep{tay92}, with
the fully integrated 2005-07-24 profiles (one for each observing
band), unless otherwise stated. Where needed, we used the timing
model derived by \citet{vbv+08} without fitting for any parameters.

\begin{table}
\centering \caption{Basic features of the selected datasets. Note
that the 2005-07-24 dataset was used only to create a timing
template and a receiver model.} \label{tab:Data}
\begin{tabular}[c]{cccccc} \\
\hline
Date       & Receiver   & Time span & Number   & File \\
           &            & (hours)   & of files & length (s) \\
\hline
2005-07-24 & MB         & 8.7  & 1596     &16.8 \\
2005-09-07 & MB         & 9.0  & 1500     &16.8 \\
2006-12-31 & H-OH       & 7.4  & 212      &67.1 \\
2007-05-06 & H-OH       & 8.9  & 152      &67.1 \\
2008-02-24 & MB         & 4.0  & 180      &67.1 \\
\hline
\end{tabular}
\end{table}

\section{Statistical tools}\label{sec:Tool}
In order to evaluate any effects on profile shape, we first
introduce the concepts of effective pulse number and pulse sharpness
below and then briefly illustrate their behaviour with pulse S/N and
TOA measurement uncertainty.

\subsection{Effective pulse number}\label{ssec:Nefc}
As pulsars are weak radio sources and individual pulses are often
not detectable, the signal needs to be folded at the rotation period
in order to obtain profiles with sufficiently high S/N to derive
precise TOAs. It is useful to check whether this procedure is as
effective as expected. Theoretically, the signal is expected to
increase linearly with integration length, while the
root-mean-square (RMS) of the noise increases according to a
square-root law. Consequently, the corresponding improvement in S/N
is expected to be proportional to the square-root of the number of
pulses. Given $N$ profiles with peak amplitudes of $A_{\rm i}$ and
noise RMSs of $\sigma_{\rm i}$ ($i=1,\ldots N$), the single pulse
S/N is:
\begin{equation}
{\rm (S/N)}_{\rm i}=\frac{A_{\rm i}}{\sigma_{\rm i}},
\end{equation}
and the S/N of a folded profile is:
\begin{equation} \label{eq:SNR_int}
{\rm S/N}=\frac{\displaystyle \sum_{i}A_{\rm
      i}}{\displaystyle \sqrt{\sum_{\rm i}\sigma_{\rm i}^{2}}}.
\end{equation}
If the profiles are identical ($A_{1} = \ldots = A_{\rm N};
\sigma_{1} = \ldots = \sigma_{\rm N}$), we have S/N $\propto
\sqrt{N}$. Practically, however, effects like intrinsic flux
variations, scintillation and system temperature variations cause
the S/Ns of profiles with identical integration times to differ.
This causes deviations from the scaling rule in the processing of
real data. Therefore, we define the effective number of pulses as:
\begin{equation}
  N_{\rm efc}=n\left(\frac{\rm S/N}{\rm S/N_{\rm mean}}\right)^{2},
\end{equation}
where $n$ is the number of pulses within an individual integration,
S/N is calculated from Eq.~(\ref{eq:SNR_int}), and $\rm S/N_{\rm
mean}$ is the averaged S/N for all integrations. Effectively,
$N_{\rm efc}$ is a normalised pulse number, which corrects for the
varying S/N of individual pulsar pulses. Consequently, the measured
S/N of averaged profiles should scale linearly with the calculated
$\sqrt{N_{\rm efc}}$, regardless of the brightness variations of the
pulses involved.

\subsection{Sharpness and calculation of TOA precision}\label{ssec:TOAcalc}
The main goals of template matching are to calculate the equivalent
TOA of the average profile with respect to a fiducial phase provided
by the template profile; and to evaluate the corresponding
uncertainty caused by the additive noise of the profile. It can be
carried out both in the time-domain and (more commonly) in the
frequency-domain, by cross-correlating the target profile with a
high S/N standard that was either obtained at a different observing
time or created analytically \citep{tay92}. Theoretically, the
uncertainty of TOA measurements, induced by the white noise of the
profile should be in the form of \citep{dr83}:
\begin{equation}\label{eq:SNRSharp}
\sigma_{\rm rn}=\frac{1}{\ss\cdot \rm
S/N_{1}}\sqrt{\frac{\Delta}{\rm N}}.
\end{equation}
Here $\rm S/N_{1}$ is the equivalent single pulse S/N and
\begin{equation}\label{eq:Sharpness}
\ss = \sqrt{\int [U'(t)]^{2}dt}
\end{equation}
is the pulse sharpness parameter, with $U(t)$ the peak-normalised
pulse waveform and
\begin{equation}
\Delta = \int\frac{\langle
n(t)n(t+\tau)\rangle}{\sigma_{n}^{2}}d\tau
\end{equation}
is the noise de-correlation time scale where $n(t)$ is the noise
function. It can be seen that the sharpness parameter basically
relates the intrinsic profile shape to the precision of the TOA.

\begin{figure}
\includegraphics[scale=0.3]{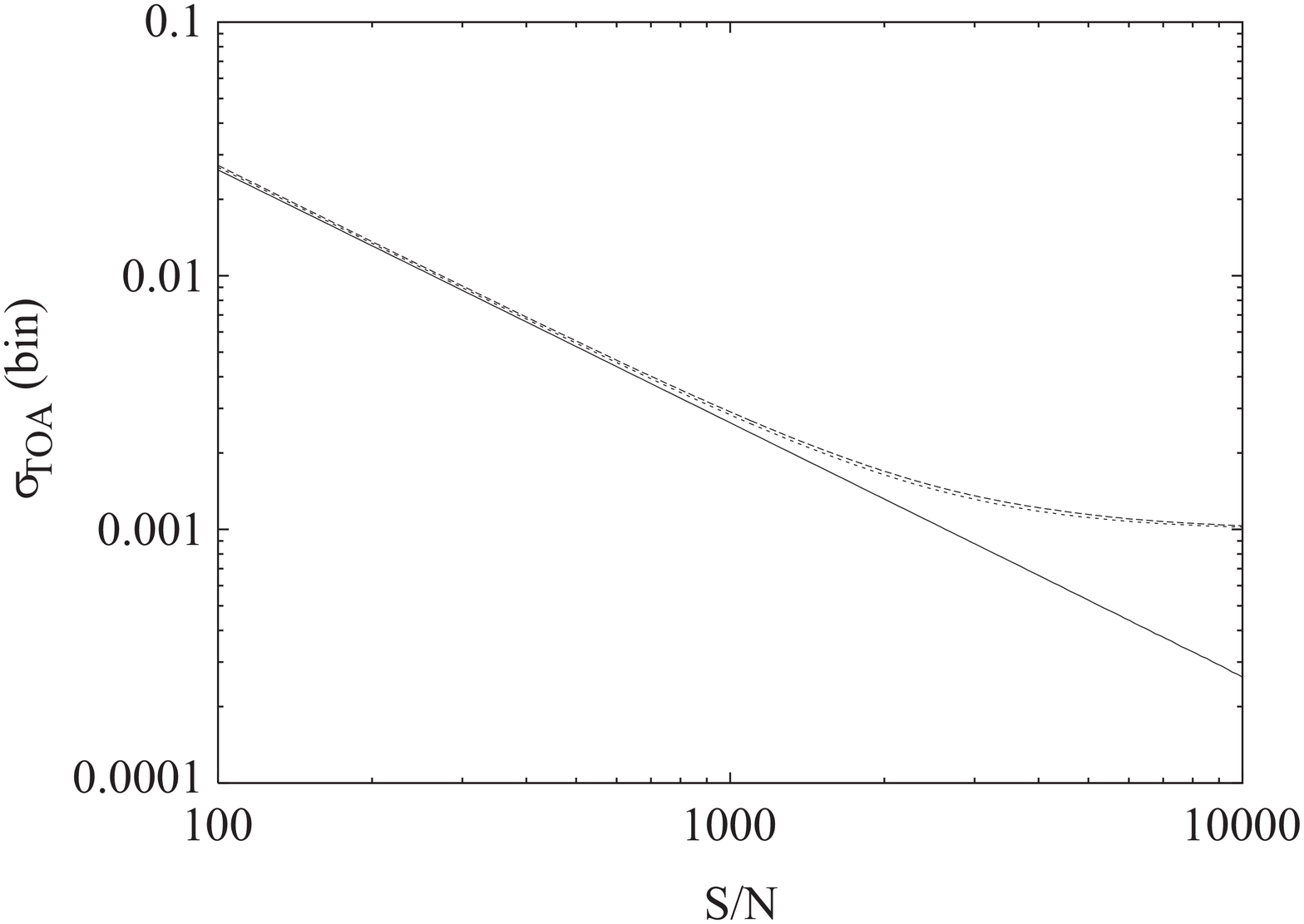}
\caption{Simulation showing the difference between ideal
  template matching (solid line) and two cases with profile distortion
  (dashed line: profile width increased by 0.5\%; dotted line: width decreased
  by 0.5\%). \label{tmpl matching simulation}}
\end{figure}

The template matching technique produces the same result as
predicted in Eq.~(\ref{eq:SNRSharp}), but only if there is no
profile shape difference between the template and the observation,
which is an ideal assumption in practice. If profile distortion (as
can be introduced by any of the effects discussed in
Section~\ref{sec:Issues}) does occur, the calculated uncertainty
does not turn out to be as good as expected for high-S/N profiles.
Fig.~\ref{tmpl matching simulation} shows a simulated example of
this case. Here a template is created as a Gaussian and fake
observation profiles are formed by adding white noise to the
template after being broadened or narrowed by $0.5\%$. It is clear
that the calculated TOA errors begin to deviate from the predicted
uncertainty once the S/N rises to values beyond 1000. Note that the
deviations are seen to be roughly equal for both the broadened and
narrowed cases, which indicates that it is not the absolute pulse
shape of the observation determining the reliability of its TOA
uncertainty, but the relative difference between the observation and
the noise-free template, stressing the importance of reliable
template profiles. The different results in the S/N$-\sigma$ graphs
presented by \cite{vbb+10} and \cite{hbb+09} already suggested this
to be the case. As their data show, even though the integrated
profile for PSR~J0437$-$4715 is supposed to be intrinsically stable
at the folding time scale, this phenomenon still appears at high
S/N.

\section{Issues affecting profile stability}\label{sec:Issues}
\begin{figure}
\includegraphics[scale=0.30]{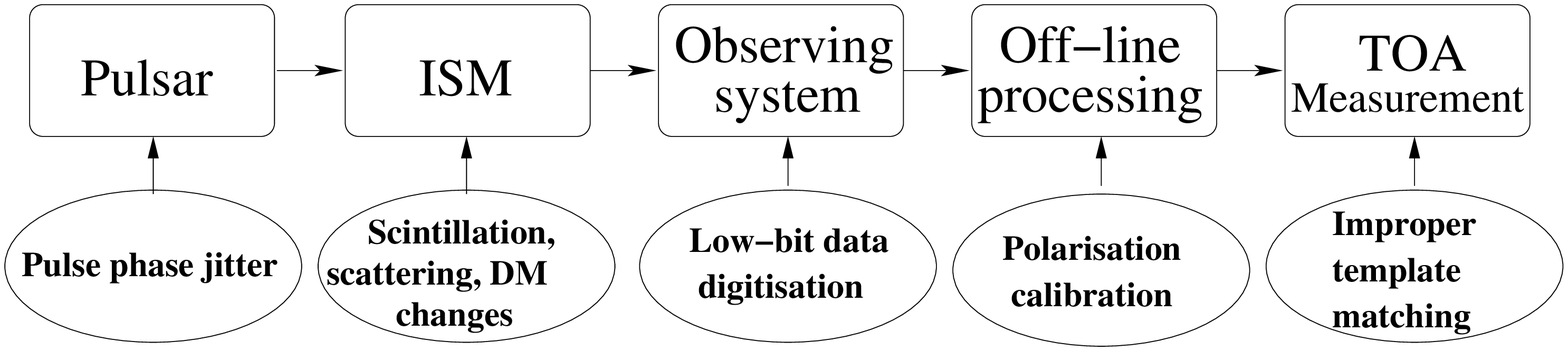}
\caption{Flow chart of the stages involved in the information
  transfer of pulsar signals. The corresponding effects that can
  lead to profile shape changes are identified for each
  stage. \label{timing}}
\end{figure}

In Fig.~\ref{timing}, we summarise the propagation path of pulsar
timing data and identify for each stage the phenomena that can
affect TOAs and their precision. Some of these are instrumental and
correctable while others induce a natural limit to timing precision.
Detailed discussions of our current knowledge about these issues
together with more in-depth investigations based on our data are
presented below.

\subsection{Dispersion changes}\label{ssec:DMvari}
\subsubsection{Theory} \label{sssec:DMtheory}
The dispersion measure (DM) is defined as the integrated free
electron density along the line of sight and introduces a delay
between two observing frequencies $\nu_1$ and $\nu_2$, in the form
of \citep[e.g.][]{lk05}
\begin{equation}
  \Delta t\simeq4.15\times10^6\rm~MHz^2pc^{-1}cm^3ms\times \left({\nu_{\rm 1}^{-2}}-{\nu_{\rm 2}^{-2}}\right)\times{\rm
  DM}. \label{eq:DM smear}
\end{equation}
There are two ways to correct for this dispersive delay in pulsar
timing data. One way is to use a filterbank to split the observing
bandwidth up into a finite number of frequency channels, which are
subsequently dedispersed with respect to each other. This is called
``incoherent dedispersion'' because the dispersion is performed post
detection when the phase information is no longer available. The
alternative approach of ``coherent dedispersion'' performs a
deconvolution on the Fourier transform of the data stream, without
loss of frequency resolution \citep{hr75}. The coherently
dedispersed data is subsequently stored with limited frequency
resolution, so any error in the DM value used in the dedispersion,
will corrupt the pulse profile permanently.

Because of turbulent motion in the ISM and the relative motions of
the pulsar, the Earth and the ISM, the integrated electron density
between the pulsar and Earth is continuously changing, implying that
ideally the DM value used in on-line coherent dedispersion would be
regularly updated to remain close to the current value. Regular
updates of the DM values are possible with (near) simultaneous
multi-frequency or wide-bandwidth observations \citep{yhc+07}, but
the accuracy of determination by this method is limited by the
system sensitivity and complicated by frequency dependent evolution
of the profile shape.

\subsubsection{Discussion}\label{sssec:DMreal}
For the PSR~J0437$-$4715 data used for this paper, given 1024 bins
across the profile and a DM of 2.644\,$\rm cm^{-3}pc$, the smearing
time (see Eq.~\ref{eq:DM smear}) within a 0.5\,MHz wide channel is
approximately 0.75 bins. Note that in reality the DM variations of
PSR~J0437$-$4715 on long timescales are below $10^{-3}$\,$\rm
cm^{-3}pc$ \citep{yhc+07}, any shape distortion induced by this
amount of DM deviation in coherent dedispersion would not be
detectable. Therefore, unless the DM variation becomes significantly
larger than previously observed, this effect does not affect our
current TOA precision. Also note that, since this type of distortion
increases the TOA uncertainty by broadening the profile, the
uncertainty will still scale following the radiometer equation for
future telescopes.

\subsection{ISM Scattering and scintillation}\label{ssec:ISM}
\subsubsection{Theory} \label{sssec:ISMtheory}
As radio pulses travel through the ionised ISM, multi-path
scattering can cause both constructive and destructive interference,
observed in the detected signal as apparent brightening and dimming
of the pulsar signal. This effect is dependent both on observing
frequency and on time because of the relative motion of the pulsar,
the Earth and the turbulent ISM. When summing frequency channels of
an observation that is affected by such scintillation, not all
frequency channels will contribute equally to the final profile, but
effectively a brightness-dependent weighting scheme will be used. In
the case where the pulse profile shape varies considerably across
the observing bandwidth, such weighting will change the resulting
pulse shape as scintles move across the observed bandwidth.

Additionally, multi-path scattering will cause different delays for
signals with different path lengths and thereby effectively broaden
the observed profile. For low DM sources observed at high frequency,
the mean of this profile broadening is non-zero but usually not
significant compared with the limited time resolution of the backend
\citep{cs10}. Still, the change of the pulse broadening function
(PBF) associated with either fast, stochastic variations in the
diffractive delay, or long-term evolution of the refraction angle,
will result in instability of the profile. The pulse broadening will
be significantly larger for high DM objects observed at low
frequency \citep{cs10}. The broadening timescale can be investigated
by assuming different scattering models, based on which it is also
possible to reveal the intrinsic profile shape by de-convolving the
pulse waveform with a theoretical broadening function
\citep{wil73,bcc+04,wksv08}.

\subsubsection{Data investigation} \label{sssec:scinreal}
The low DM of PSR~J0437$-$4715 (2.644\,cm$^{-3}$pc) corresponds to a
short broadening timescale $\tau_{\rm s}$ ($<10^{-5}$\,ms) and a
large decorrelation frequency scale \citep[estimated by $\Delta
f\sim1/(2\pi\tau_{\rm s})\approx0.4\,$GHz, see ][]{cl01,bcc+04}.
Consequently, provided the effective 48\,MHz bandwidth\footnote{As
described in Section \ref{sec:Obs}, the total bandwidth is 64\,MHz
but on either side of the bandpass 12.5\,\% was removed, leaving
48\,MHz of effective bandwidth.} of our data, the shape change on
short timescales by the PBF variation should be negligible here.

\begin{figure}
\includegraphics[scale=0.3]{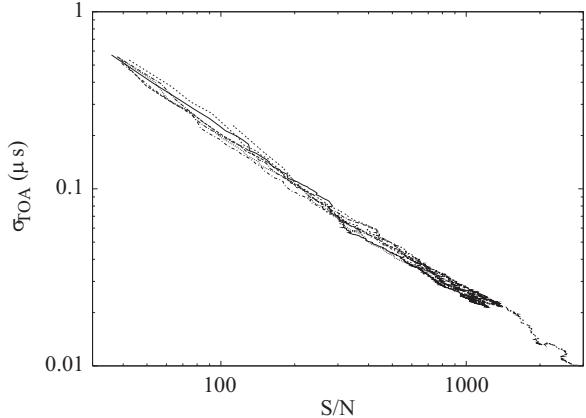}
\caption{Sub-band template matching for the 2005-09-07 dataset at
  1405\,MHz. Curves correspond to different bands and the double dashed line
  is for the entire bandwidth combined. The sub-band curves yield a set
  of TOAs which are equivalent to a single
  TOA obtained from the full-band curve.
  \label{snrsigmamultifreq}}
\end{figure}

The influence on pulse shape, by the combination of scintillation
and profile shape evolution over frequency, can be investigated by
dividing the entire bandwidth into sub-bands and carrying out
individual template matching. The templates for the sub-bands are
produced with the same bandwidth and central frequency as the
averaged profiles. Fig.~\ref{snrsigmamultifreq} presents an example
from the 2005-09-07 dataset. Here the effective bandwidth of 48\,MHz
is divided into six bands of 8\,MHz each. The S/N is calculated from
off-pulse bins which show Gaussian statistics. The plot shows that
all of the matching processes have identically behaved
S/N$-\sigma_{\rm TOA}$ curves and the same final $\sigma_{\rm TOA}$
of about 22\,ns. Statistically, the six TOAs with an uncertainty of
22\,ns each, are equivalent to a single one of uncertainty
$22/\sqrt{6}\simeq9$\,ns. This is the TOA precision we obtain from
template matching the fully frequency-integrated observation. It
means that across the 48\,MHz bandwidth of this dataset, the TOA
precision does not benefit from conducting sub-band template
matching. The effect could become significant when the observing
bandwidth is comparable to or larger than the scintillation
frequency scale. Still, as the effect can be dealt with through the
application of frequency-dependent template matching, we conclude
that it will not be a limiting factor to TOA precision with either
current or future telescopes.

\subsection{Signal digitisation effects} \label{ssec:digitisation}
\subsubsection{Phenomena} \label{sssec:digittheory}
Shape distortion induced by instrumental effects obscures the true
pulse shape and can therefore be expected to decrease the precision
of timing. Two main digitisation effects are pertinent to a low-bit
observing system.

The first effect is caused by the underestimation of the undigitised
power in a system with low dynamic range (e.g. only 2 bits per
sample). As discussed in \cite{ja98} (hereafter JA98) the earlier
arrival of pulsar emission at the high end of the observing
bandwidth causes an increase in undigitised power, and therefore a
decrease of the digitised-to-undigitised power ratio at all
frequencies if the output power level is kept constant. As a result,
the off-pulse power will be decreased in the rest of the band. This
effect can be avoided through dynamically setting the output power
levels, which provides the required dynamic output range and
therefore does not result in negative off-pulse dips on either side
of the pulse profile.

The second artefact is caused by quantisation errors as a
second-order distortion, and manifests itself as an increase in
white noise uniformly redistributed across all frequency channels,
which is induced by the increase in pulsed power in one part of the
band (JA98). This scattered power broadens the profile and causes
additional pulse shape variations as a function of observing
frequency, which decreases the achievable TOA precision.

\subsubsection{Correction}\label{sssec:digitreal}
\begin{figure*}
\includegraphics[scale=0.3]{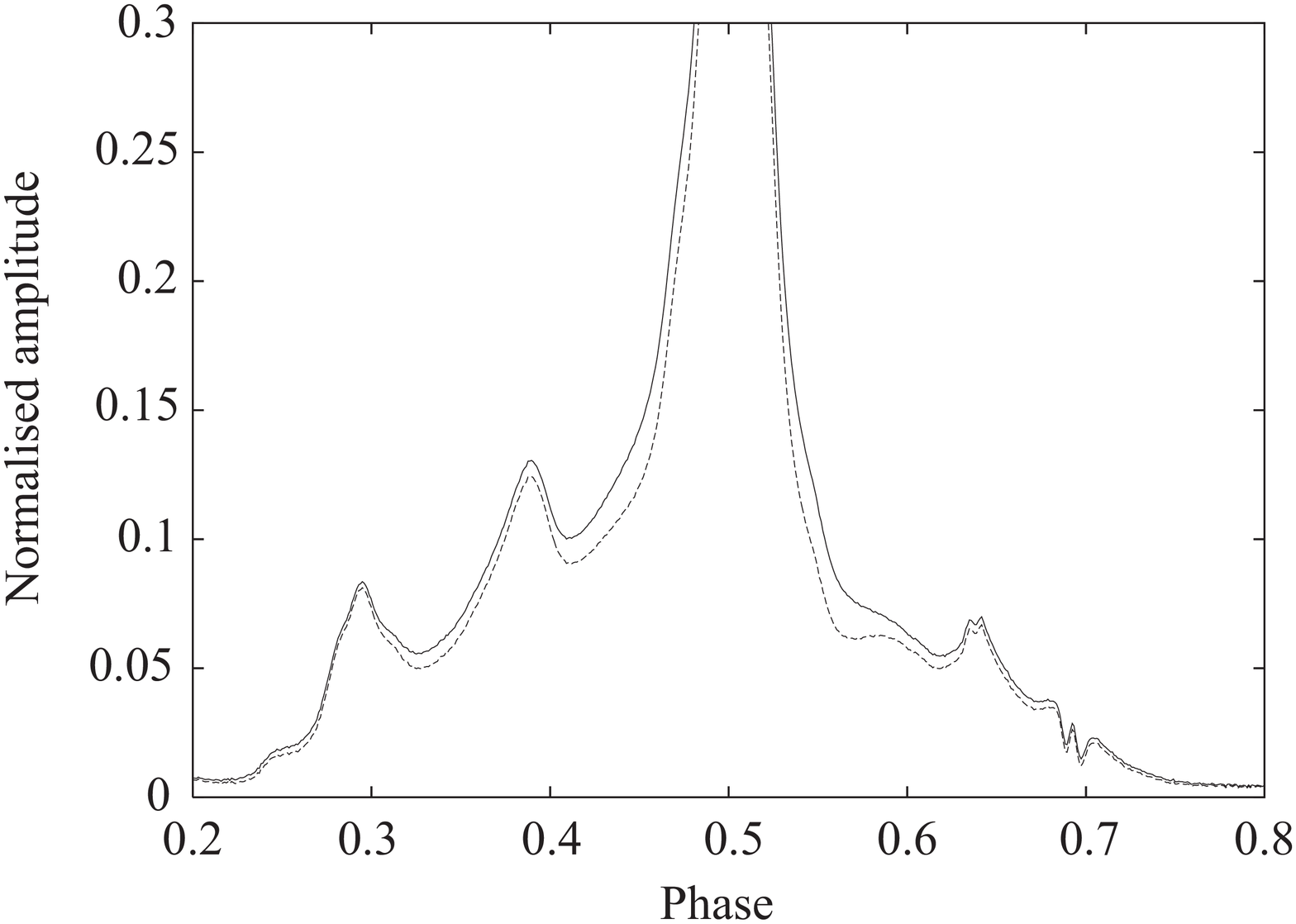}
\includegraphics[scale=0.3]{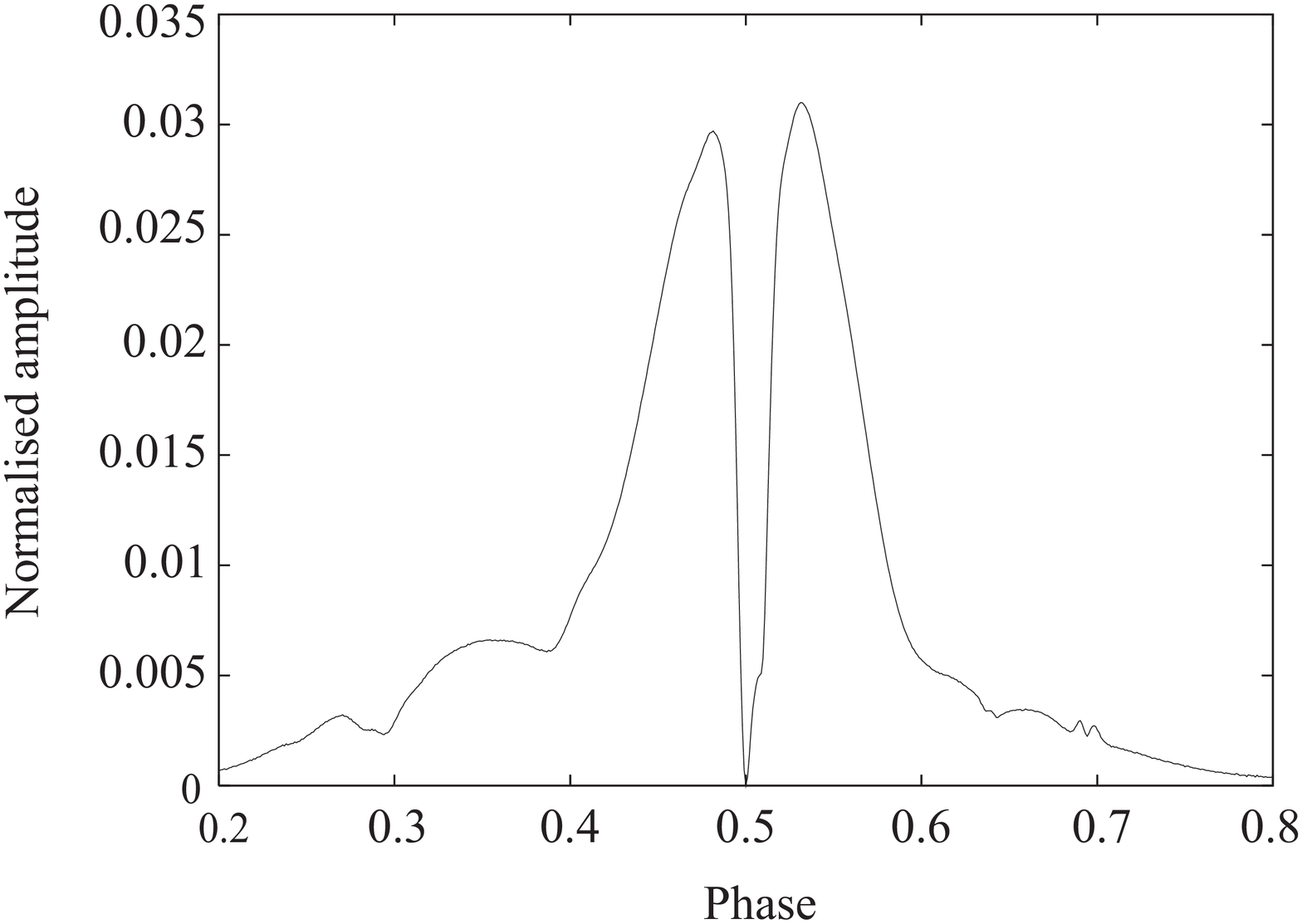}
\caption{Pulse profiles at 1405\,MHz formed from the 2005-07-24
dataset, demonstrating the effect of digitisation scattered power on
pulse shape. The left-hand plot shows the zoomed-in pulse profile
before (solid line) and after (dashed line) application of the
correction algorithm, while the right-hand plot presents the
corresponding difference in this phase range, clearly showing the
broader shape of the uncorrected profile. \label{profresispc}}
\end{figure*}

In this paper, all CPSR2 data presented were corrected for the low
dynamic range artefact during on-line processing, by the dynamic
output level setting algorithm implemented in {\sc
dspsr}\footnote{http://dspsr.sourceforge.net/} \citep{vb10}. The
scattered power was mitigated during off-line processing through
application of the correction algorithm implemented in {\sc
psrchive} \citep{hvm04}. Given the uncorrected, mean digitized power
$\hat\sigma^2$ in each pulse phase bin, this algorithm inverts
Eq.~(A5) of JA98 to estimate the mean undigitised power $\sigma^2$
and the mean scattered power $A$ via Eqs.~(45) and (43) of JA98,
respectively. The effect of correction is demonstrated in
Fig.~\ref{profresispc}, which shows the pulse profile formed from
the 2005-07-24 dataset with and without application of the
algorithm, as well as the difference between the two. The decreased
pulse width of the corrected profile allows higher timing precision.
Note that the distortion would not change significantly once the
back-end settings are stable. This means that the TOA precision
would still scale with effective collecting area as described by the
radiometer equation. We therefore conclude that this effect does not
limit the current TOA precision, and will not limit it for future
telescopes either, which are likely to employ digitisers with a
higher number of digitisation levels.

\subsection{Polarimetric calibration imperfections}\label{ssec:Calib}
\subsubsection{Theory} \label{sssec:calibtheory}
When a fixed-linear-feed, alt-azimuth radio telescope tracks a
polarised source across the sky, the feed will rotate with respect
to the plane of polarisation by the parallactic angle ($q$), which
is defined as the angle between the object-zenith great circle and
the hour circle. The change of $q$ combined with the instrumental
response, will result in a variation of the observed Stokes
parameters with time. Polarisation calibration, the aim of which is
to reveal the intrinsic profile, will correct this time-dependent
variation, but the correction will only be partial if any
non-orthogonality of the receptors is not fully modelled. In this
case, a difference between profiles at different $q$ will be seen
even after calibration. In practice, a ``single axis'' model
considering only the differentials in gain and phase for the two
linear polarisation probes is usually applied
\citep[e.g.][]{scr+84}. The most recent (here mentioned as ``full
reception'') model, described by \cite{van04a}, solves the matrix
description of the polarisation measurement equations, accounting
for differential gains and phases, as well as for coupling and
leakage effects between the receiver feeds.

\subsubsection{Calibration} \label{sssec:polcalmod}
\begin{figure*}
  \mbox{ \subfigure[]{
      \includegraphics[scale=0.28]{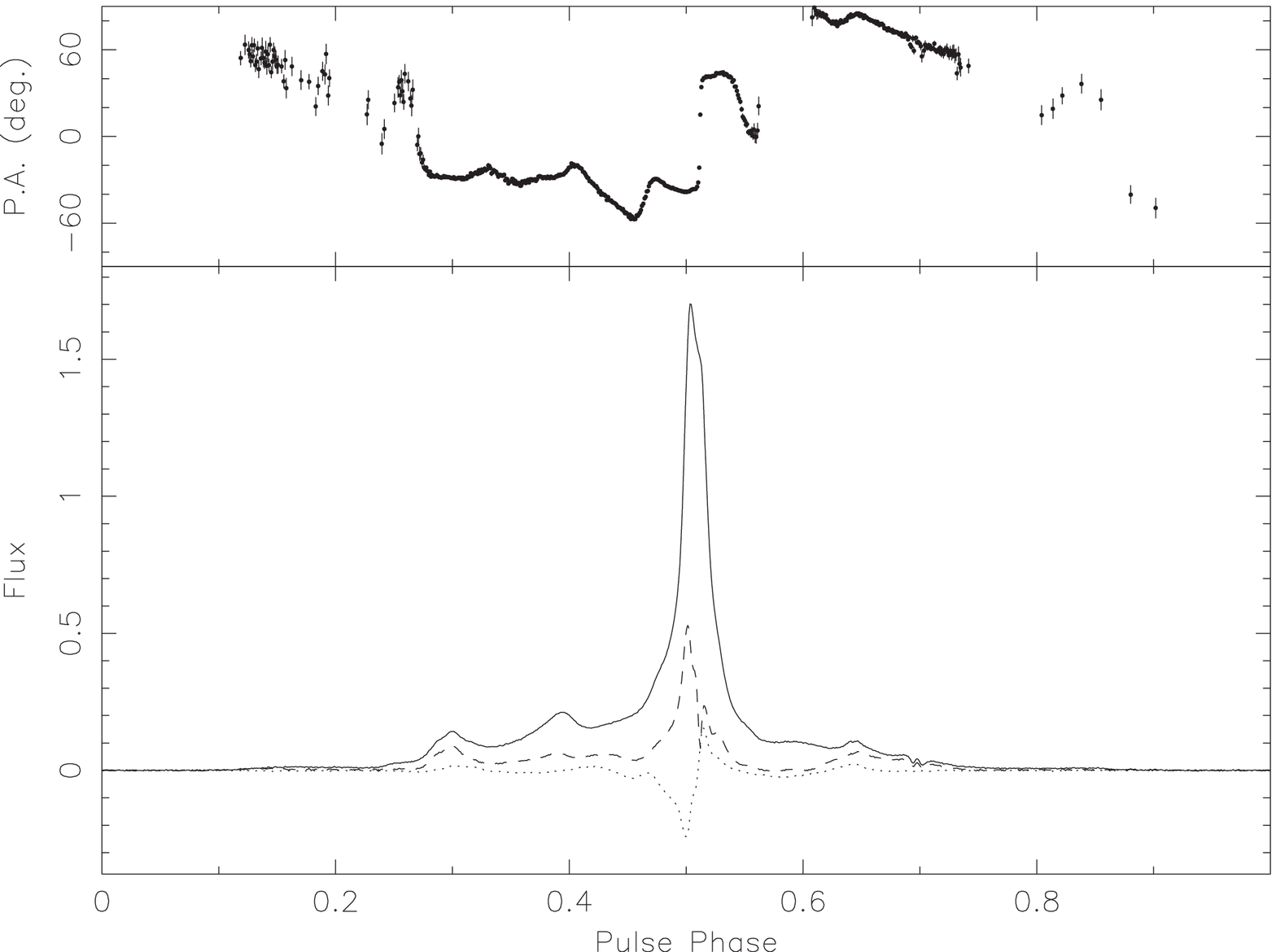} }
    \subfigure[]{
      \includegraphics[scale=0.28]{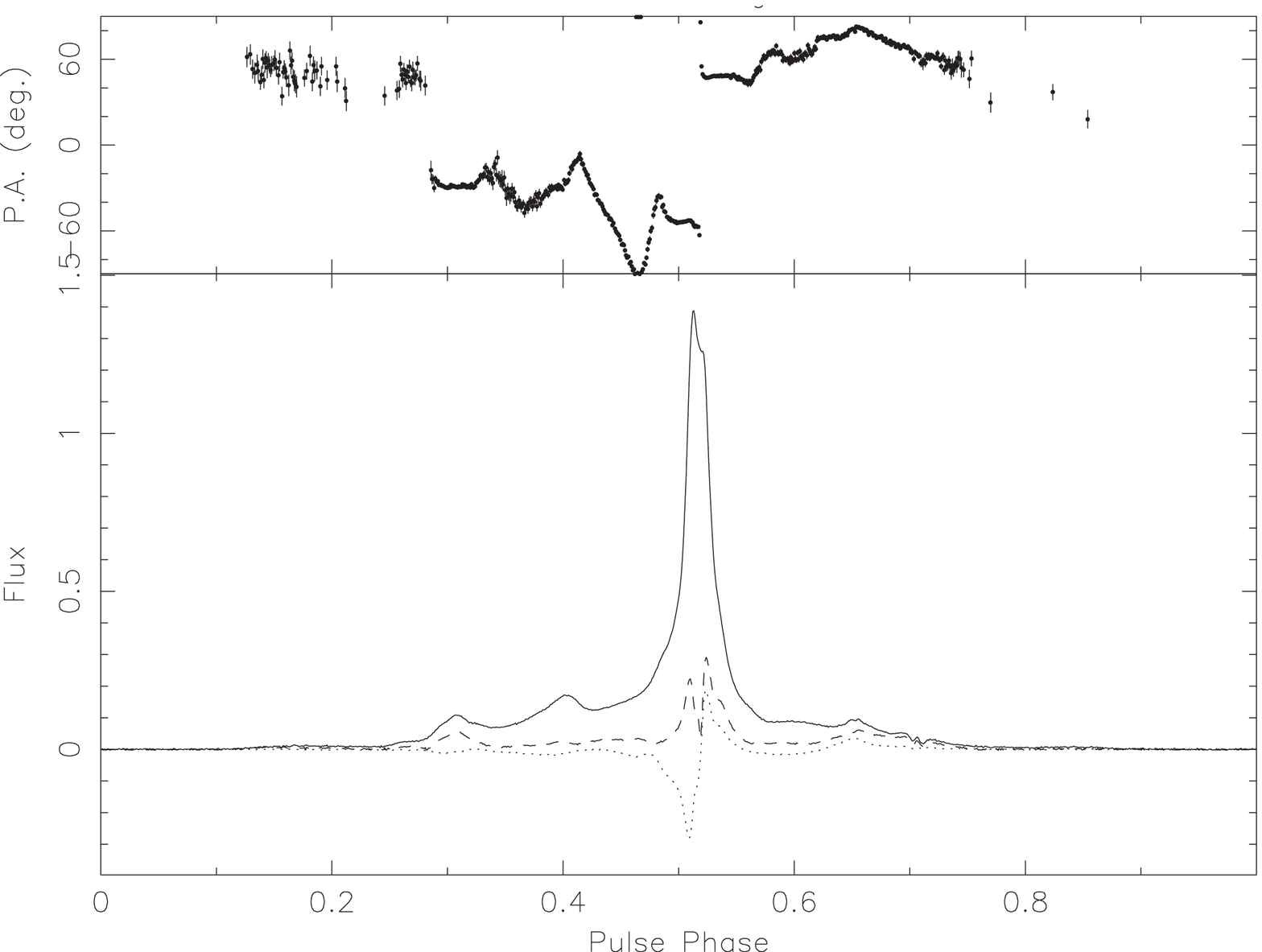} }
    \subfigure[]{
      \includegraphics[scale=0.28]{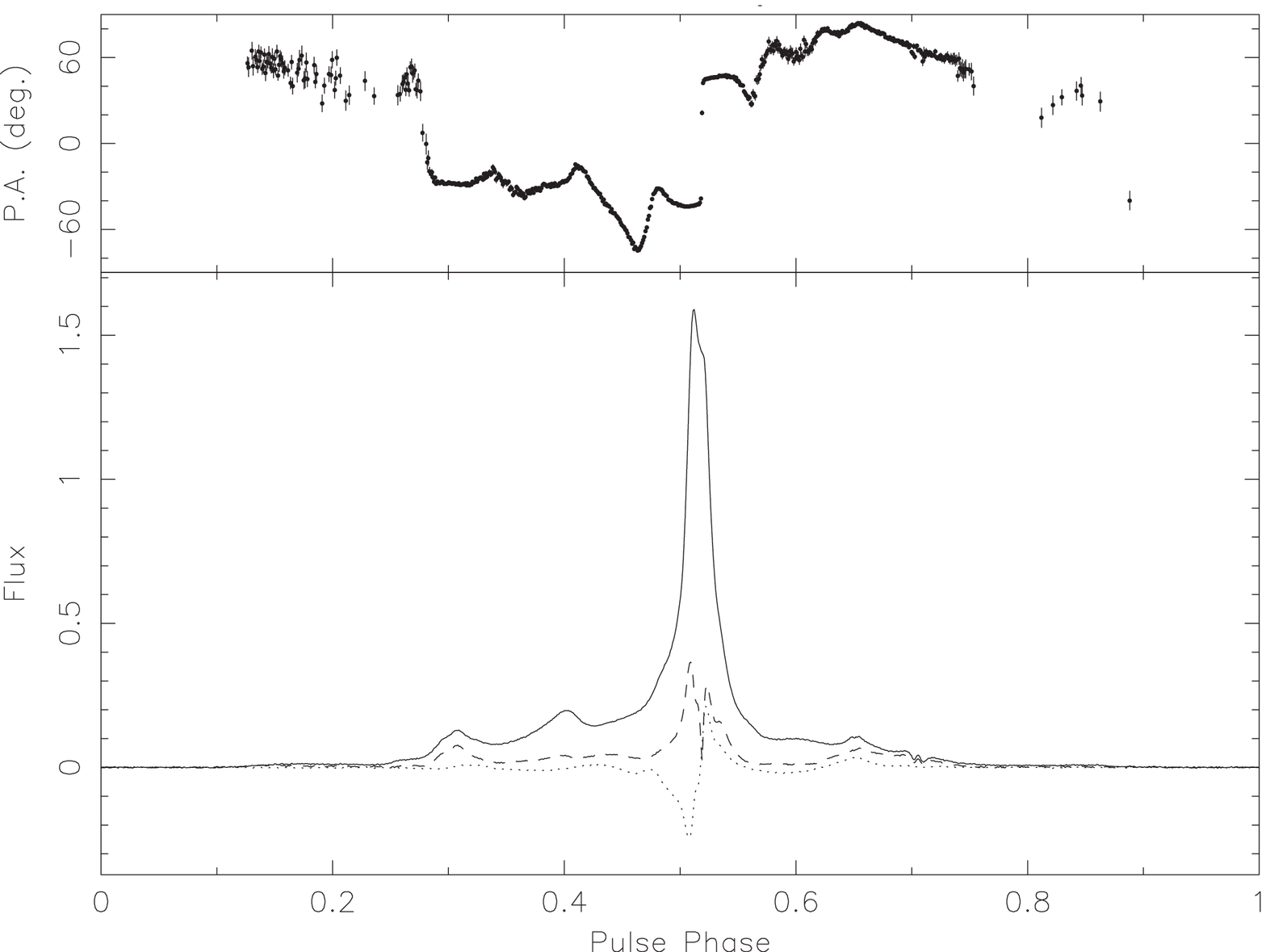} }
  } \mbox{ \subfigure[]{
      \includegraphics[scale=0.5, angle=-90]{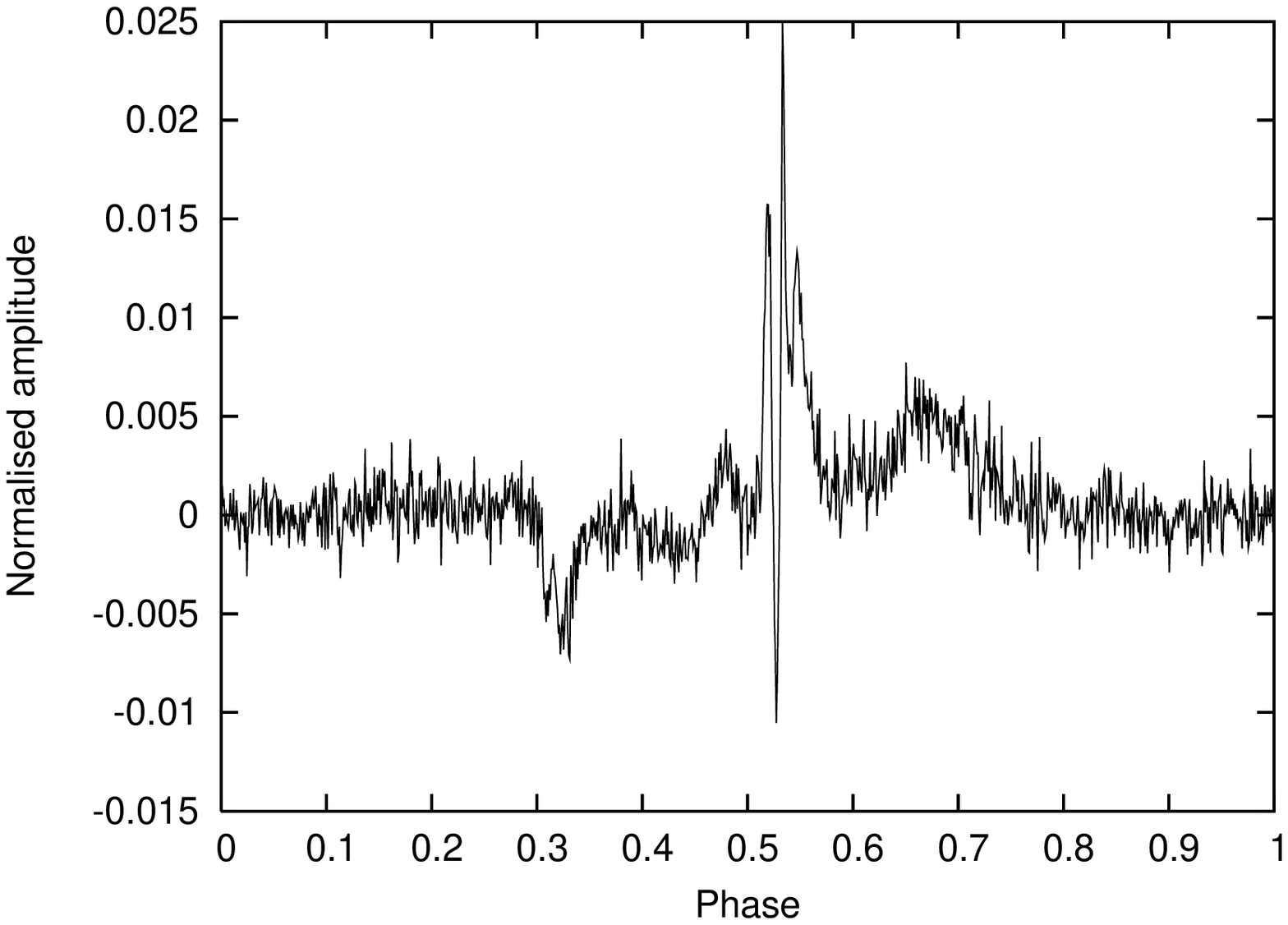}}
    \subfigure[]{
      \includegraphics[scale=0.5, angle=-90]{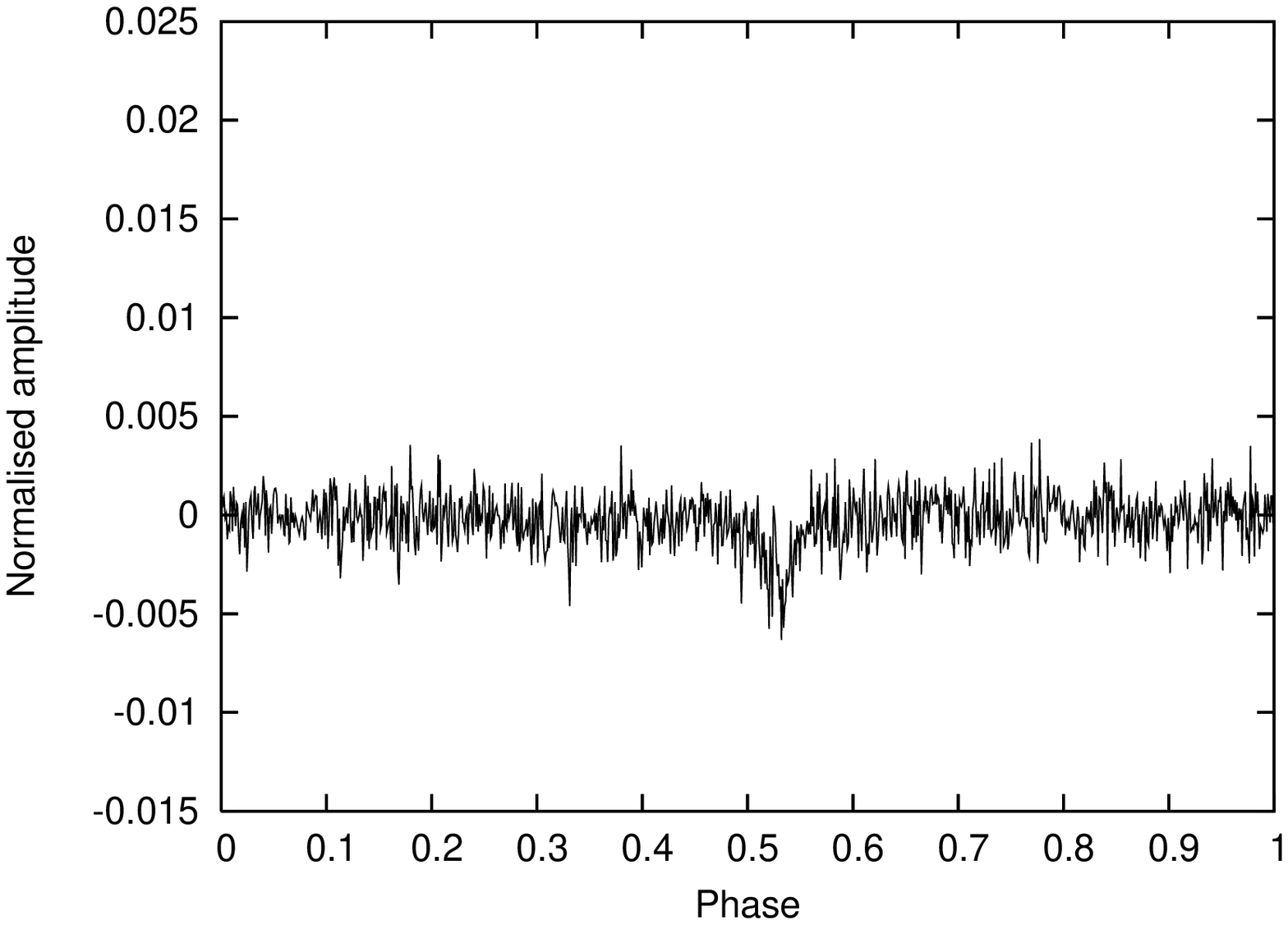}} }
  \caption{Comparison of polarimetric calibration models. Shown are
    two hour-long observations from the 2005-09-07 dataset at
    1405\,MHz at different parallactic angles and calibrated with the
    single axis (subplots a and b) and full reception models (subplot
    c, identical for both hours of observations). Here the solid,
    dashed, and dotted line correspond to the total intensity, linear
    polarisation, and circular polarisation profiles, individually.
    The difference of the total intensity profiles for the two hours of observations are
    shown in subplot (d) for the single-axis calibration scheme and in
    subplot (e) for the full reception model calibration.
    \label{polaresi}}
\end{figure*}

For comparison, here we used both the ``single axis'' and the ``full
reception'' model (constructed from the 2005-07-24 dataset) for
calibration of the 2005-09-07 dataset. Fig.~\ref{polaresi} (a)-(b)
shows two hour-long integrated profiles formed from the 2005-09-07
dataset, covering a different range of parallactic angles and
calibrated according to the single axis model. The large differences
in linear polarisation and position angle demonstrate the
imperfection of this calibration model. The difference between the
total intensity profile at the two different values of $q$ is shown
in subplot (d) of Fig.~\ref{polaresi}. The same profile as in
subplot (a), but calibrated with the full reception model, is shown
in subplot (c) - this profile is identical for both observing times,
as the difference plot (e) shows. Furthermore, the profiles
calibrated with the full reception model compare well with the
previously published polarimetry of \cite{nms+97}, regardless of
$q$. The remaining difference is clearly far less than 2\% of the
total intensity, and is therefore below the uncertainty level of the
calibration, as quantified through simulations by \cite{ovhb04}.
Further simulation shows that the profile difference in plot (e)
will induce TOA errors of less than 30\,ns. It is therefore clear
that the full reception model removes all polarisation calibration
artefacts down to the level of our current TOA precision on
PSR~J0437$-$4715. We hence conclude that polarimetric calibration
does not limit the current TOA accuracy above $~30$\,ns. Though the
application of these calibration schemes to future interferometers
(such as SKA) remains to be solved, such accuracy will be achievable
for future single-dish telescopes (such as FAST).

\subsection{Interference and unknown observing system instabilities}\label{ssec:other effect}
\subsubsection{Diagnostic theory} \label{sssec:2bittheory}
Besides the expected effects described above, there are some
unpredictable effects that may also affect the data quality. Radio
frequency interference (RFI) and instrumental failure are the most
important two. Specifically in the case of an observing system with
only four digitisation levels (i.e. a 2-bit system), any excess in
power (as potentially caused by RFI) or a temporary non-linear
response in the system, can be expected to affect the pulse shape.

In order to keep track of any such occurrences, the statistics of
the digitised data can be compared to those expected from theory, in
the following way. First, the digitised data are divided into
consecutive segments of $L$ samples and, for each segment, the
number of low-voltage states $M$ is counted. The digital signal
processing software that is used to process the 2-bit data,
maintains a histogram of occurrences of $M$ that is archived with
the pulsar data for later use as a diagnostic tool \citep{vb10}.
When the voltage input to the digitiser is normally distributed, the
ratio $\Phi = M/L$ has a binomial distribution as in Eq.~(A6) of
\citet{ja98}. The difference between this theoretical expectation
and the recorded histogram of $M$ provides a measure of 1) the
degree to which the input signal deviates from a normal
distribution, and 2) the degree to which the sampling thresholds
diverge from optimality. This difference, called the {\it 2-bit
distortion}, is given by
\begin{equation}\label{eq:2bitStat}
  D = \sum_{M=0}^L \left[{\mathcal P}(M/L) - {\mathcal H}(M) \right]^2
\end{equation}
where ${\mathcal P}(\Phi)$ is the expected binomial distribution and
${\mathcal H}(M)$ is the recorded distribution of $M$. Separate
histograms of $M$ are maintained for each polarisation, and the
reported distortion is simply the sum of the distortion in each
polarisation.

\subsubsection{Data analysis}\label{ssec:2bitdist}
\begin{figure*}
    \includegraphics[scale=0.6]{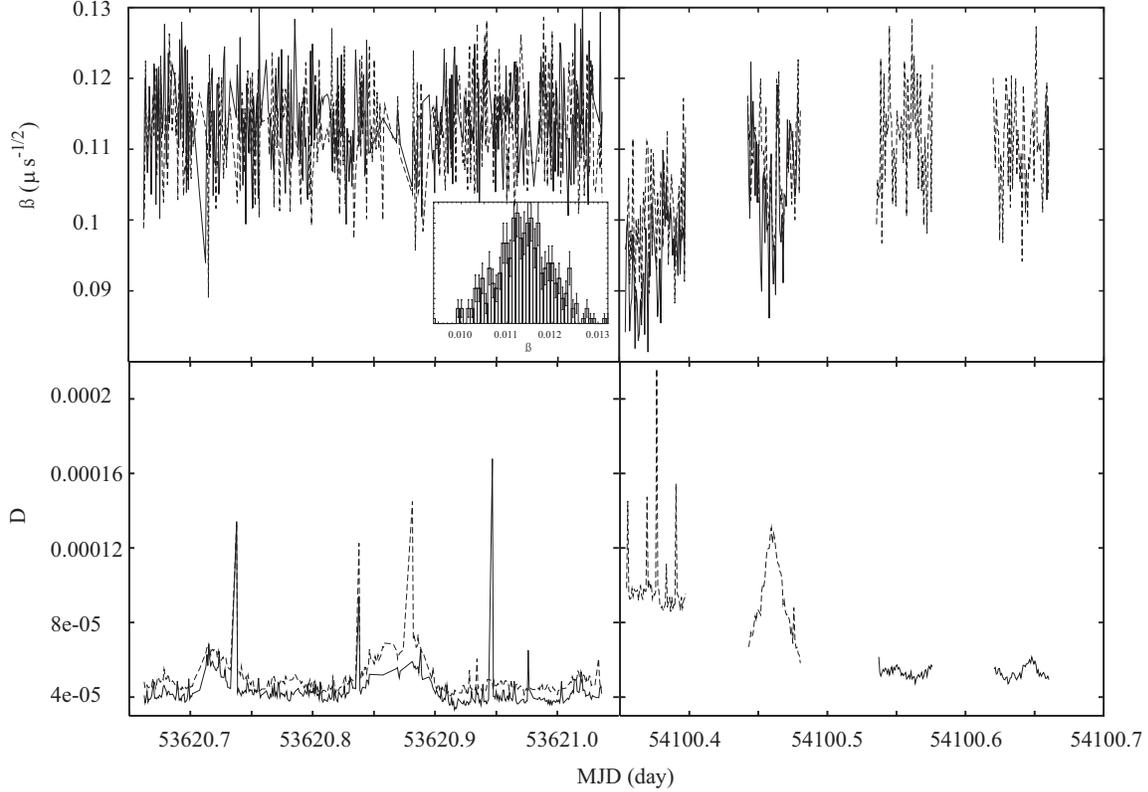}
  \caption{Sharpness $\ss$ (top) and 2-bit distortion $D$ (bottom) for
    the observations of 2005-09-07 (left-hand plots) and 2006-12-31
    (right-hand plots). Results of the observing band centred at
    1405\,MHz are shown with solid lines; those of the 1341\,MHz
    observing band with dashed lines. The shape variations on
    2006-12-31 identified by the changes in sharpness are clearly
    correlated with changes in 2-bit distortion, suggesting these
    variations are instrumental rather than intrinsic to the pulsar.
    The sharpness variations of the 2005-09-07 data are close to
    Gaussian as their histogram (inset) shows.
    \label{tsharp t2bit}}
\end{figure*}

The sharpness parameter can be used to examine profile stability
over short lengths of observing time. In Fig.~\ref{tsharp t2bit},
the variations of sharpness and 2-bit distortion (defined in
Eq.~\ref{eq:2bitStat}) during the observing runs of 2005-09-07 and
2006-12-31 are shown side by side. Note that the S/N of the
individual integrations is typically not sufficient to apply
Eq.~(\ref{eq:Sharpness}) directly. The sharpness is therefore
indirectly obtained through Eq.~(\ref{eq:SNRSharp}). In doing so,
the white noise uncertainty is estimated from cross-correlation with
the single integration and the S/N is measured by calculation of the
off-peak RMS. As the profile has a large on-pulse duty cycle, only
$\sim15\%$ of the pulse period can be used to calculate the noise
RMS, which results in a $\sim10\%$ uncertainty of both the measured
S/N and the sharpness estimates. The 2-bit distortion is calculated
by following the stages laid out in Section~\ref{sssec:2bittheory}.

As shown in Fig.~\ref{tsharp t2bit} the 2-bit distortion levels were
consistently low during the 2005-09-07 observation and the sharpness
was close to normally distributed. The observation of 2006-12-31,
however, shows significantly larger levels of 2-bit distortion as
well as correlated, non-Gaussian variations in the sharpness
parameter. Furthermore, the 2-bit distortion has a high degree of
correlation with the sharpness levels. The correlation coefficient
of the two time series is $-0.73$, suggesting that the profile
distortion is related to the digitisation.

\begin{figure}
  \includegraphics[scale=0.3]{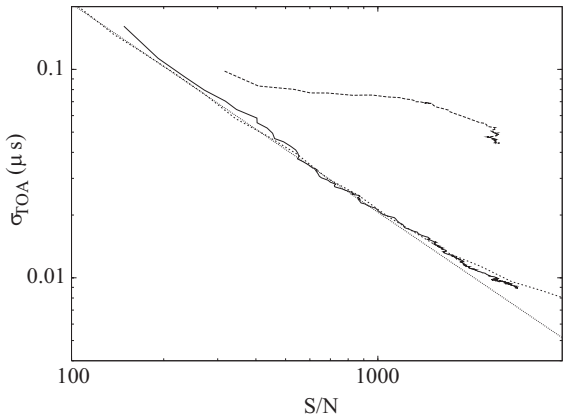}
  \caption{S/N-$\sigma_{\rm TOA}$ relations for both real data
    and simulations. The solid line represents the result
    from the 2005-09-07 dataset, the long-dashed line from the
    2006-12-31 dataset. The dotted line shows the theoretical
    prediction from Eq.~(\ref{eq:SNRSharp}), and the short-dashed line
    shows the relationship for simulated profiles cross-correlated
    with a noisy standard. All these curves are for the observing band
    centred at 1405\,MHz.\label{snrsigmas}}
\end{figure}

To illustrate the effect these digitisation-induced shape changes
may have on TOA precision, we provide the ${\rm S/N}-\sigma_{\rm
TOA}$ plot for the data analysed in Fig.~\ref{snrsigmas}. Clearly
the 2005-09-07 data behave as expected: they follow the theoretical
inverse relationship and worsen slightly for S/N$>1000$. This
worsening is caused by noise in the template profile, which was
constructed from the 2005-07-24 dataset. To demonstrate this, we
created the ${\rm S/N}-\sigma_{\rm TOA}$ curve for simulated data,
based on a (simulated) template profile with a S/N identical to that
of the 2005-07-24 standard profile. This simulated result is shown
as the dotted line in Fig.~\ref{snrsigmas} and follows the
2005-09-07 curve well. Ideally, therefore, a noise-free analytic
template profile would be used \citep[as in][]{kxc+99}, but the
small-scale features present in the profile of PSR~J0437$-$4715
require advanced modelling that goes beyond the scope of this paper.
This may also explain the flattening of the $\sigma_{\rm TOA}$-S/N
curve in \cite{vbb+10}.

The ${\rm S/N}-\sigma_{\rm TOA}$ curve for the 2006-12-31 data
displays much larger deviations: for equal S/N its TOA uncertainty
($\sigma_{\rm TOA}$) is several factors higher than for the
2005-09-07 data and limits the calculated precision to the $\sim
50$\,ns level.

The other datasets in our investigation (2005-07-24, 2007-05-06 and
2008-02-24) also yield well-behaved S/N$-\sigma_{\rm TOA}$ curves
like that of 2005-09-07, which shows that the majority of our data
is not affected by this digitisation artefact. It must be noted that
the only objective statistic that we found to clearly identify the
2006-12-31 data as corrupted, was the 2-bit distortion statistic
defined by Eq.~(\ref{eq:2bitStat}) and that this corruption cannot
be corrected post-detection. This demonstrates the need to
investigate the digitisation distortion in any 2-bit pulsar timing
data and to exclude any distorted data from future pulsar timing
analyses. We conclude that these statistics should be used to
identify and discard corrupted data. We also note that for systems
with more digitisation levels, this type of distortion will not
affect TOA precision as much (or at all). It is therefore likely
that timing with more state-of-the-art systems on both present and
future telescopes, will not be limited by these effects.

\subsection{Pulse jitter}\label{ssec:pulsejitter}
\subsubsection{Theoretical model} \label{sssec:jitter theory}
The phases of single pulses vary around an expected, average phase.
For some pulsars these phase variations seem random \citep{em68} and
are called ``jitter'', while for others there are clearly systematic
drifts called ``drifting sub-pulses'' \citep{sspw70}. These phase
variations slightly broaden the pulse shape and induce additional
arrival phase fluctuation of integrated profiles. For the case of
pulse jitter, assuming no modulation of the single pulse shape and
intensity, the jitter-induced TOA scatter can be written as
\citep{cs10}:
\begin{equation}
\sigma_{\rm J}=\left[\frac{f^{2}_{\rm J}\int U(t)t^{2}dt}{N\int
U(t)dt}\right]^{1/2},\label{eq:jitter}
\end{equation}
where $N$ is the number of integrated pulses (assuming no systematic
drifting), $f_{\rm J}$ is the width of the probability density
function (PDF) of the phase jitter in units of pulse width, and
$U(t)$ is the normalised pulse waveform. Note that this uncertainty
is attributed to emission stability of the source and is not related
to the observing hardware. One can see that the higher the system
sensitivity is, the more important pulse phase jitter may become.

\subsubsection{Data analysis}\label{sssec:realjitter}
\begin{figure}
\includegraphics[scale=0.3]{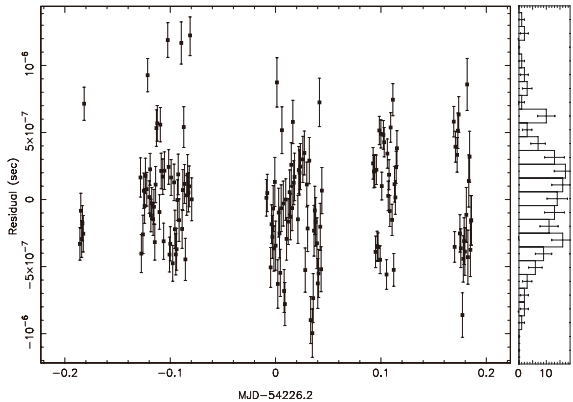}
\caption{Timing residuals of the 2007-05-06 dataset from the
1405\,MHz side-band data. The TOAs derived from 1-min integrations
are clearly scattered on a scale larger than the estimated errors
with a reduced chi-square of 8.1. The right column shows the
histogram of the residuals. \label{timresi}}
\end{figure}

Pulse jitter can be studied by a few methods and in the following we
describe each approach in detail.
\paragraph{Timing rms} \label{par:jtiming}
The first way to evaluate the impact of pulse phase jitter on
timing, is to study the random variations of the TOAs after a timing
model has been subtracted. When we consider the timing residuals for
the 2007-05-06 dataset, we notice that the TOAs are widely scattered
and the reduced chi-square is well above unity (shown in
Fig.~\ref{timresi}). This is so because the TOA uncertainties have
been determined based purely on the amount of radiometer noise in
the pulse profile, while other possible contributions of uncertainty
such as phase jitter, timing model imperfections, short-term
interstellar instabilities and instrumental effects remain
unquantified. Note that a previous single pulse study has shown no
evidence for pulse drifting \citep{jak+98}. The dataset also has
passed through the 2-bit distortion test (mentioned in
Section~\ref{ssec:2bitdist}) and the residuals satisfy a
distribution close to Gaussian, which suggests the insignificance of
effects such as faulty ephemeris and improper calibration that can
induce non-white noise in timing. The deviation from Gaussian
distribution is attributed to both small number statistics, and to
differences in measurement precision of the residuals caused by S/N
variations in individual integrations. Consequently, we can set up
an upper limit on the phase jitter contribution, by assuming that
jitter noise is dominant in the additional residual scatter (the
amount not quantified by the TOA uncertainties from radiometre
noise) and using Eq.~(\ref{eq:jitter}) to derive the parameter
$f_{\rm J}$. We find that in this worst-case scenario, $f_{\rm
J}\simeq 0.08$ and $\sigma_{\rm Rad}/\sigma_{\rm J}\simeq 0.3$,
which means that the timing residuals are dominated by pulse phase
jitter and can therefore hardly be improved by increasing the
telescope gain, but only by extending the integration
time\footnote{The result is consistent for the other datasets and a
detailed study will be provided in a subsequent paper}.

\begin{figure}
\includegraphics[scale=0.3]{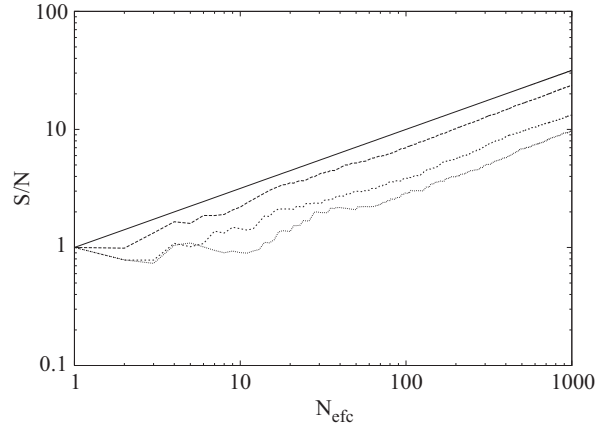}
\caption{Simulated $N_{\rm efc}$-S/N curves for both the jitter free
and jittered profile cases. The solid, long dashed, short dashed,
and dotted lines correspond to jitter factors $f_{\rm J}$ (see Eq.
2) of 0, 0.1, 0.3 and 0.5, respectively. \label{Nefcsimu}}
\end{figure}

\begin{figure*}
\includegraphics[scale=0.4]{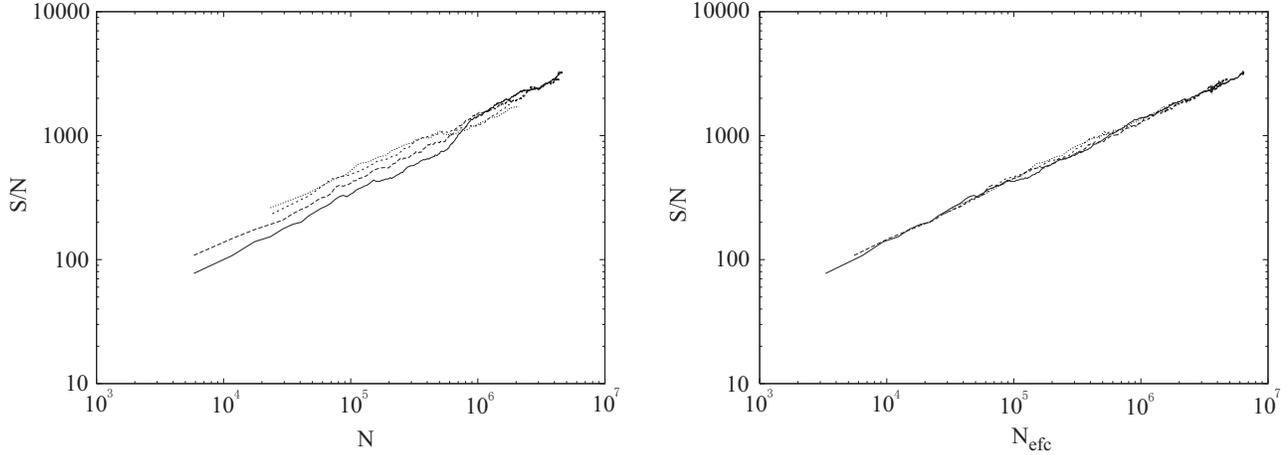}
\caption{Real and effective pulse number versus S/N, demonstrating
the
  relations between integration time and S/N. These plots are based on
  data at 1405\,MHz from 2005-07-24 (solid line), 2005-09-07 (long
  dashed line), 2007-05-06 (short dashed line) and 2008-02-24 (dotted
  line). The theoretical scaling law is clearly reproduced in the
  $N_{\rm efc}$ graph but not in the true $N$ graph.\label{NefcSNR}}
\end{figure*}

\paragraph{$N_{\rm efc}$-S/N relation} \label{par:jnefc}
A second approach to investigate pulse jitter is through the $N_{\rm
efc}-$S/N relation introduced in Section~\ref{ssec:Nefc}. A
deviation from its theoretical scaling can be induced by inaccurate
folding, non-Gaussian noise in the off-pulse baseline or, more
interestingly, by pulse jitter. Based on the model of
Eq.~(\ref{eq:jitter}), the effect of pulse phase jitter on the
$\sqrt{N_{\rm efc}}$-S/N scaling law has been evaluated through
Monte-Carlo simulations, as shown in Fig.~\ref{Nefcsimu}. Here the
template of PSR~J0437$-$4715 is used as a single pulse shape, and
for each pulse we apply a Gaussianly distributed shift based on a
given $f_{\rm J}$ value. Clearly, the inclusion of pulse phase
jitter introduces an initial deviation that is strongly dependent on
the value of $f_{\rm J}$. After integration of a sufficiently large
number of pulses (of the order of a few to several tens of pulses),
the scaling law is recovered while the curves remain at a lower S/N
than in the jitter-free case. Fig.~\ref{NefcSNR} complements
Fig.~\ref{Nefcsimu} by plotting the same parameters over a range of
increasing integration times. While Fig.~\ref{Nefcsimu} shows the
simulated effects of pulse phase jitter at the very shortest
integration lengths that are inaccessible to our data (which have a
minimum length of 16.8 seconds or just under 3000 pulses),
Fig.~\ref{NefcSNR} shows the (expectedly stable) behaviour on longer
timescales. The $\rm S/N_{\rm ref}$ used for the calculation of
$N_{\rm efc}$ is 150, which is roughly the mean value for all
67.1\,s integrations. The left plot underlines how the S/N does not
scale linearly with the real number of pulses averaged while the
right plot does follow the theoretical linear relationship as
expected. The phenomenon shown is visible only when the shortest
integration time can resolve of the order of ten pulses. These plots
do not show any evidence of the effects simulated in
Fig.~\ref{Nefcsimu}. This is to be expected, as the simulations show
that the jitter effect saturates after a relatively short
integration time, which is not accessible by the (relatively long)
integrations of the data we have.


\begin{figure}
\includegraphics[scale=0.3]{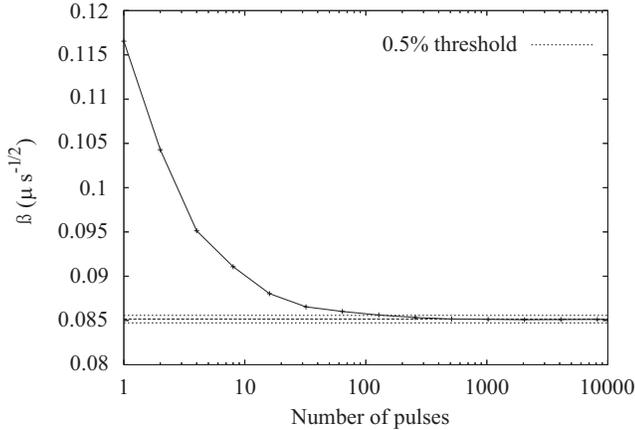}
\caption{Simulated integration process in the existence of pulse
jitter. The profile shape estimated by the sharpness parameter is
well constrained below the 0.5 \% level after folding of a few
hundred pulses. Note that given the rotational period of
PSR~J0437$-$4715, $10^{4}$ pulses roughly correspond to an
integration time of 1 minute. \label{fig:Nsharp}}
\end{figure}

\paragraph{Sharpness variation} \label{par:jsharp}
A third diagnostic plot that can be used to analyse pulse phase
jitter, is the $\ss-N_{\rm efc}$ plot. An example of this is
provided in Fig.~\ref{fig:Nsharp}, which shows how $\ss$ converges
as more pulses are integrated. This simulation (which was performed
in a similar way to that described above) is for the extreme
jitter-dominated case. Real data would be a combination of the
jitter-induced exponential convergence simulated in
Fig.~\ref{fig:Nsharp} and of Gaussian white noise. However, because
the intrinsic shape of single pulses for PSR~J0437$-$4715 is not
easily defined, we cannot compare Fig.~\ref{fig:Nsharp} easily with
the $\ss$ values obtained from our data (as shown in
Fig.~\ref{tsharp t2bit}), but we do point out that $\ss$ does not
change significantly as we integrate our data. Furthermore, the
variation of $\ss$ shown in Fig.~\ref{tsharp t2bit} (top left panel)
is far larger than the variation suggested by the simulations
(Fig.~\ref{fig:Nsharp}). This suggests either that the simulated
level of pulse phase jitter is well below the radiometer noise
present in the data, or that there are significant variations in the
shape of single pulses, where our model assumes there are none.

\subsubsection{Conclusion} \label{sssec:jconclu}
In summary, we have used timing measurements to derive the upper
limit of $f_{\rm J} \leq 0.08$ for pulse phase jitter in
PSR~J0437$-$4715. This upper limit was subsequently used in
simulations to evaluate the impact such jitter would have on the
pulse shape and we found that at integration times beyond 100 pulses
($\sim 0.57$\,s) neither the S/N of the integrated profile, nor the
pulse sharpness are affected significantly compared to the
radiometer noise. This lack of pulse jitter effects on the S/N and
sharpness of the simulated data, implies either that the derived
upper limit is overly conservative and that the underestimation of
TOA uncertainties is mostly induced by causes other than pulse phase
jitter, or that our model for pulse phase jitter is overly
simplistic and a more complex model is required. A more detailed
investigation is required to distinguish between these two scenarios
and to quantify the pulse phase jitter in PSR~J0437$-$4715 more
precisely. Such an analysis is being prepared for a subsequent paper
\citep{lkl+11}.

\section{Discussion and consequences for future telescopes}
\label{sec:Conclusions}
\subsection{Summary}\label{ssec:sum}
In this paper, we have used the brightest and most precisely timed
MSP, PSR~J0437$-$4715, to illustrate the most important phenomena
known to affect the shape of pulse profiles; to estimate the impact
of these phenomena and to evaluate the efficacy of mitigation
schemes. By concentrating on the brightest MSP currently known, we
are able to cast some light on effects that future telescopes like
FAST or the SKA will come across as a matter of course in any
standard MSP.

We find that pulse phase jitter may be dominant in current
short-term timing of PSR~J0437$-$4715, though further analysis is
required to enable quantification of this effect. The effects of
faulty de-dispersion are found to be small, negating the need for
frequent updates of the dedispersion DM. The effects of
scintillation across the bandwidth could not be fully investigated
given the limited bandwidth of our data, but usage of
frequency-dependent templates can be expected to resolve the
aforementioned problem that scintillation might cause. We present
the results of the application of correction algorithms for the most
important digitisation artefacts and illustrate the importance of
full calibration modelling, as opposed to the more traditional
calibration for differential gain and phase only. Finally, we point
out the importance of adequate monitoring of the digitisation
statistics for 2-bit pulsar observing systems, which is not commonly
practised.

We further propose a few diagnostic plots to assess the data quality
of any pulsar timing data:
\begin{itemize}
\item Time-$\ss$ curve: Any non-Gaussian variations in profile
  sharpness suggests changes in pulse shape and imply the data
  should be carefully studied before being included in any timing
  analysis (e.g. Fig.~\ref{tsharp t2bit}).
\item S/N-$\sigma_{\rm TOA}$ relation: To assess the quality of the
  standard profile used (Fig.~\ref{tmpl matching simulation}). Deviations from the theoretical relationship
  or significant differences between datasets indicate data
  distortion (e.g. Fig.~\ref{snrsigmas}).
\item S/N-$\sigma_{\rm TOA}$ for sub-bands: As in the previous point,
  any mismatch between curves of different sub-bands indicates loss of
  data quality (e.g. Fig.~\ref{snrsigmamultifreq}).
\end{itemize}
Note that the analysis is applicable not only to future telescopes,
but also to existing ones such as Parkes, Arecibo, Effelsberg, and
the Large European Array for Pulsars \citep[LEAP, see][]{fvb+10}, if
the system sensitivity is high enough given the brightness of the
source.

\subsection{Timing in the new millennium}
Since PSR~J0437$-$4715 is two orders of magnitude brighter than most
MSPs, current observations of this pulsar provide a good
demonstration of the TOA measurement situation for most of the MSPs
by the next generation of radio telescopes. The 21$^{\rm st}$
century radio telescope will significantly increase the S/N of
pulsar detections and correspondingly reduce the uncertainty of TOA
measurements through vast increases in effective collecting area.

\begin{figure}
  \includegraphics[scale=0.45]{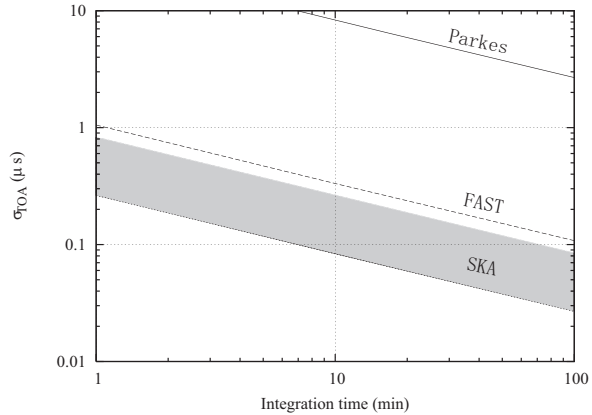}
  \caption{Expected average-brightness MSP TOA scatter versus integration time for
    different instruments. Only radiometer noise and pulse
    jitter are accounted for as influences on the measurement accuracy.
    \label{sigmasimu}}
\end{figure}

Fig.~\ref{sigmasimu} shows the expected MSP TOA precision that can
be achieved by SKA and FAST. Here for Parkes and FAST we only
consider the uncertainty induced by radiometer noise and pulse
jitter at the worst-case level derived in
Section~\ref{sssec:realjitter}, while for the prediction of SKA we
calculate the upper limit by applying the worst-case in jitter and
set up the lower limit by assuming no pulse jitter at all. The
instrumental effects causing profile distortions and phase
fluctuations are neglected based on previous analyses and the ISM
influence on pulse shape is assumed to be either corrected
\citep[e.g.][]{wksv08} or not significant on a short timescale with
provided frequency and bandwidth. Concerning the properties of MSPs,
we assume a mean flux density of 3.0\,mJy at 1.4\,GHz, 50 MHz
bandwidth, spin period of 5.0\,ms and 5\% pulse width. In addition,
a 10\,K sky temperature at 1.4\,GHz is applied. Consequently, it is
shown that the TOA precision for normal MSPs can be improved by over
one order of magnitude with the next generation of radio telescopes,
even when assuming a worst-case level for pulse phase jitter. The
result indicates that jitter induced uncertainties will be
considerable in future timing of MSPs. The instrumental parameters
used for these calculations are presented in Table~\ref{tab:FutTel}
\citep{sac+07,nan06}.

We conclude that at 1.4\,GHz for 10-min integrations, a TOA
precision of between 80 and 230\,ns can be expected in timing of
normal brightness MSPs by SKA. The wide range of this prediction is
caused by our limited knowledge of the pulse jitter mechanism. The
future radio telescopes will enable deeper investigation of profile
stability of single pulses. Presently the comparatively high levels
of radiometer noise mean that only a few bright MSPs have allowed a
systematic study of their single pulses, where only a subset of the
single pulses can be clearly detected \citep{cstt96,jak+98,es03}.
The significant improvement in instrumental sensitivity will both
dramatically increase the number of MSPs available for single pulse
analysis and reduce the selection effects present in current
studies.

\begin{table}
  \centering \caption{Key parameters used in prediction of TOA
    precision for different instruments at 1.4\,GHz.}\label{tab:FutTel}
  \begin{tabular}[c]{ccc} \\
    \hline
    &$A_{\rm eff}$ ($\rm m^{2}$)  &$T_{\rm sys}$ (K)\\
    \hline
    Parkes  &$2.2\times10^{3}$       &23.5   \\
    FAST    &$7.1\times10^{4}$       &20.0   \\
    SKA core&$2.5\times10^{5}$       &30.0   \\
    \hline
  \end{tabular}
\end{table}

\subsection{Limitations and future work}
The analysis based on PSR~J0437$-$4715 is limited by the selected
sample. Note that the DM and scattering timescale (see
Section~\ref{ssec:ISM}) of this pulsar are relatively low so that
the influence of the ISM on the TOA precision within the provided
bandwidth is expected to be negligible. As system sensitivity and
observational bandwidth are improved, more high-DM sources will be
timed at high precision TOA measurements. The effects of scattering
and scintillation on profile shape will then become more
considerable than shown in this paper, and will need to be accounted
for by using correction algorithms \citep{bcc+04,wksv08} and/or
frequency-dependent template profiles.

\section*{Acknowledgments}
We thank A.~G.~Lyne and M.~Purver for their advice and help on this
research. The Parkes telescope is part of the Australia Telescope
which is funded by the Commonwealth of Australia for operation as a
National Facility managed by CSIRO. KL is funded by a stipend of the
Max-Planck-Institute for Radio Astronomy. JPWV is supported by the
European Union under Marie Curie Intra-European Fellowship 236394.

\bibliographystyle{mn}
\bibliography{journals,psrrefs,modrefs,crossrefs}

\begin{thebibliography}{43}
\expandafter\ifx\csname natexlab\endcsname\relax\def\natexlab#1{#1}\fi

\bibitem[{{Bhat} {et~al.}(2004){Bhat}, {Cordes}, {Camilo}, {Nice}, \&
  {Lorimer}}]{bcc+04}
{Bhat} N.~D.~R., {Cordes} J.~M., {Camilo} F., {Nice} D.~J., {Lorimer} D.~R.,
  2004, 605, 759

\bibitem[{Cognard {et~al.}(1996)Cognard, Shrauner, Taylor, \&
  Thorsett}]{cstt96}
Cognard I., Shrauner J.~A., Taylor J.~H., Thorsett S.~E., 1996, 457, L81

\bibitem[{{Cordes} {et~al.}(2004){Cordes}, {Kramer}, {Lazio}, {Stappers},
  {Backer}, \& {Johnston}}]{ckl+04}
{Cordes} J.~M., {Kramer} M., {Lazio} T.~J.~W., {Stappers} B.~W., {Backer}
  D.~C., {Johnston} S., 2004, The North American Review, 48, 1413

\bibitem[{{Cordes} \& {Lazio}(2001)}]{cl01}
{Cordes} J.~M., {Lazio} T.~J.~W., 2001, 549, 997

\bibitem[{{Cordes} \& {Shannon}(2010)}]{cs10}
{Cordes} J.~M., {Shannon} R.~M., 2010, arXiv:1010.3785

\bibitem[{Downs \& Reichley(1983)}]{dr83}
Downs G.~S., Reichley P.~E., 1983, 53, 169

\bibitem[{{Edwards} \& {Stappers}(2003)}]{es03}
{Edwards} R.~T., {Stappers} B.~W., 2003, Acta Arithmetica, 407, 273

\bibitem[{{Ekers} \& {Moffet}(1968)}]{em68}
{Ekers} R.~D., {Moffet} A.~T., 1968, 220, 756

\bibitem[{{Ferdman} {et~al.}(2010){Ferdman}, {van Haasteren}, {Bassa},
  {Burgay}, {Cognard}, {Corongiu}, {D'Amico}, {Desvignes}, {Hessels},
  {Janssen}, {Jessner}, {Jordan}, {Karuppusamy}, {Keane}, {Kramer},
  {Lazaridis}, {Levin}, {Lyne}, {Pilia}, {Possenti}, {Purver}, {Stappers},
  {Sanidas}, {Smits}, \& {Theureau}}]{fvb+10}
{Ferdman} R.~D., {van Haasteren} R., {Bassa} C.~G., {Burgay} M., {Cognard} I.,
  {Corongiu} A., {D'Amico} N., {Desvignes} G., {Hessels} J.~W.~T., {Janssen}
  G.~H., {Jessner} A., {Jordan} C., {Karuppusamy} R., {Keane} E.~F., {Kramer}
  M., {Lazaridis} K., {Levin} Y., {Lyne} A.~G., {Pilia} M., {Possenti} A.,
  {Purver} M., {Stappers} B., {Sanidas} S., {Smits} R., {Theureau} G., 2010,
  Classical and Quantum Gravity, 27, 084014

\bibitem[{{Hankins} \& {Rickett}(1975)}]{hr75}
{Hankins} T.~H., {Rickett} B.~J., 1975, in Methods in Computational Physics
  Volume 14 --- Radio Astronomy, Academic Press, New York, pp. 55--129

\bibitem[{{Hobbs} {et~al.}(2010){Hobbs}, {Coles}, {Manchester}, \&
  {Chen}}]{hcmc10}
{Hobbs} G., {Coles} W., {Manchester} R., {Chen} D., 2010, arXiv:1011.5285

\bibitem[{{Hobbs} {et~al.}(2009){Hobbs}, {Jenet}, {Lee}, {Verbiest}, {Yardley},
  {Manchester}, {Lommen}, {Coles}, {Edwards}, \& {Shettigara}}]{hjl+09}
{Hobbs} G., {Jenet} F., {Lee} K.~J., {Verbiest} J.~P.~W., {Yardley} D.,
  {Manchester} R., {Lommen} A., {Coles} W., {Edwards} R., {Shettigara} C.,
  2009, 394, 1945

\bibitem[{Hobbs {et~al.}(2009)Hobbs, Bailes, Bhat, Burke-Spolaor, Champion,
  Coles, Hotan, Jenet, Kedziora-Chudczer, Khoo, Lee, Lommen, Manchester,
  Reynolds, Sarkissian, van Straten, To, Verbiest, Yardley, \& You}]{hbb+09}
Hobbs G.~B., Bailes M., Bhat N.~D.~R., Burke-Spolaor S., Champion D.~J., Coles
  W., Hotan A., Jenet F., Kedziora-Chudczer L., Khoo J., Lee K.~J., Lommen A.,
  Manchester R.~N., Reynolds J., Sarkissian J., van Straten W., To S., Verbiest
  J.~P.~W., Yardley D., You X.~P., 2009, 26, 103

\bibitem[{Hotan {et~al.}(2006)Hotan, Bailes, \& Ord}]{hbo06}
Hotan A.~W., Bailes M., Ord S.~M., 2006, 369, 1502

\bibitem[{{Hotan} {et~al.}(2004){Hotan}, {van Straten}, \&
  {Manchester}}]{hvm04}
{Hotan} A.~W., {van Straten} W., {Manchester} R.~N., 2004, 21, 302

\bibitem[{Jenet {et~al.}(1998)Jenet, Anderson, Kaspi, Prince, \&
  Unwin}]{jak+98}
Jenet F., Anderson S., Kaspi V., Prince T., Unwin S., 1998, 498, 365

\bibitem[{Jenet \& Anderson(1998)}]{ja98}
Jenet F.~A., Anderson S.~B., 1998, 110, 1467

\bibitem[{{Jenet} {et~al.}(2006){Jenet}, {Hobbs}, {van Straten}, {Manchester},
  {Bailes}, {Verbiest}, {Edwards}, {Hotan}, \& {Sarkissian}}]{jhv+06}
{Jenet} F.~A., {Hobbs} G.~B., {van Straten} W., {Manchester} R.~N., {Bailes}
  M., {Verbiest} J.~P.~W., {Edwards} R.~T., {Hotan} A.~W., {Sarkissian} J.~M.,
  2006, 653, 1571

\bibitem[{Johnston {et~al.}(1993)Johnston, Lorimer, Harrison, Bailes, Lyne,
  Bell, Kaspi, Manchester, D'Amico, Nicastro, \& Jin}]{jlh+93}
Johnston S., Lorimer D.~R., Harrison P.~A., Bailes M., Lyne A.~G., Bell J.~F.,
  Kaspi V.~M., Manchester R.~N., D'Amico N., Nicastro L., Jin S., 1993, Neues
  Archiv der Gesellschaft f{\"u}r {\"a}ltere deutsche Geschichtskunde, 361, 613

\bibitem[{Kramer {et~al.}(2004)Kramer, Backer, Cordes, Lazio, Stappers, \&
  Johnston}]{kbc+04}
Kramer M., Backer D.~C., Cordes J.~M., Lazio T. J.~W., Stappers B.~W., Johnston
  S., 2004, The North American Review, 48, 993

\bibitem[{Kramer {et~al.}(2006)Kramer, Stairs, Manchester, McLaughlin, Lyne,
  Ferdman, Burgay, Lorimer, Possenti, D'Amico, Sarkissian, Hobbs, Reynolds,
  Freire, \& Camilo}]{ksm+06}
Kramer M., Stairs I.~H., Manchester R.~N., McLaughlin M.~A., Lyne A.~G.,
  Ferdman R.~D., Burgay M., Lorimer D.~R., Possenti A., D'Amico N., Sarkissian
  J.~M., Hobbs G.~B., Reynolds J.~E., Freire P. C.~C., Camilo F., 2006, 314, 97

\bibitem[{Kramer {et~al.}(1999)Kramer, Xilouris, Camilo, Nice, Lange, Backer,
  \& Doroshenko}]{kxc+99}
Kramer M., Xilouris K.~M., Camilo F., Nice D., Lange C., Backer D.~C.,
  Doroshenko O., 1999, 520, 324

\bibitem[{Lee {et~al.}(2008)Lee, Jenet, \& Price}]{ljp08}
Lee K.~J., Jenet F.~A., Price R.~H., 2008, 685, 1304

\bibitem[{{Liu} {et~al.}(2011){Liu}, {Keane}, {Lee}, \& {Kramer}}]{lkl+11}
{Liu} K., {Keane} E.~F., {Lee} K.~J., {Kramer} M., 2011, in preparation

\bibitem[{Lorimer \& Kramer(2005)}]{lk05}
Lorimer D.~R., Kramer M., 2005, {Handbook of Pulsar Astronomy}. Cambridge
  University Press

\bibitem[{{Nan}(2006)}]{nan06}
{Nan} R., 2006, Science in China G: Physics and Astronomy, 49, 129

\bibitem[{Navarro {et~al.}(1997)Navarro, Manchester, Sandhu, Kulkarni, \&
  Bailes}]{nms+97}
Navarro J., Manchester R.~N., Sandhu J.~S., Kulkarni S.~R., Bailes M., 1997,
  486, 1019

\bibitem[{{Ord} {et~al.}(2004){Ord}, {van Straten}, {Hotan}, \&
  {Bailes}}]{ovhb04}
{Ord} S.~M., {van Straten} W., {Hotan} A.~W., {Bailes} M., 2004, 352, 804

\bibitem[{{Schilizzi} {et~al.}(2007){Schilizzi}, {Alexander}, {Cordes},
  {Dewdney}, {Ekers}, {Faulkner}, {Gaensler}, {Hall}, {Jonas}, \&
  {Kellerman}}]{sac+07}
{Schilizzi} R.~T., {Alexander} P., {Cordes} J.~M., {Dewdney} P.~E., {Ekers}
  R.~D., {Faulkner} A.~J., {Gaensler} B.~M., {Hall} P.~J., {Jonas} J.~L.,
  {Kellerman} K.~I., 2007, Preliminary specifications for the square kilometre
  array. Memo 100, SKA Program Development Office

\bibitem[{Staveley-Smith {et~al.}(1996)Staveley-Smith, Wilson, Bird, Disney,
  Ekers, Freeman, Haynes, Sinclair, Vaile, Webster, \& Wright}]{swb+96}
Staveley-Smith L., Wilson W.~E., Bird T.~S., Disney M.~J., Ekers R.~D., Freeman
  K.~C., Haynes R.~F., Sinclair M.~W., Vaile R.~A., Webster R.~L., Wright
  A.~E., 1996, 13, 243

\bibitem[{Stinebring {et~al.}(1984)Stinebring, Cordes, Rankin, Weisberg, \&
  Boriakoff}]{scr+84}
Stinebring D.~R., Cordes J.~M., Rankin J.~M., Weisberg J.~M., Boriakoff V.,
  1984, 55, 247

\bibitem[{Sutton {et~al.}(1970)Sutton, Staelin, R.M., \& Weimer}]{sspw70}
Sutton J.~M., Staelin D.~H., R.M. P., Weimer R., 1970, 159, L89

\bibitem[{Taylor(1992)}]{tay92}
Taylor J.~H., 1992, 341, 117

\bibitem[{Taylor \& Weisberg(1989)}]{tw89}
Taylor J.~H., Weisberg J.~M., 1989, 345, 434

\bibitem[{van Straten(2004)}]{van04a}
van Straten W., 2004, 152, 129

\bibitem[{{van Straten} \& {Bailes}(2011)}]{vb10}
{van Straten} W., {Bailes} M., 2011, 28, 1

\bibitem[{van Straten {et~al.}(2001)van Straten, Bailes, Britton, Kulkarni,
  Anderson, Manchester, \& Sarkissian}]{vbb+01}
van Straten W., Bailes M., Britton M., Kulkarni S.~R., Anderson S.~B.,
  Manchester R.~N., Sarkissian J., 2001, Neues Archiv der Gesellschaft f{\"u}r
  {\"a}ltere deutsche Geschichtskunde, 412, 158

\bibitem[{{Verbiest} {et~al.}(2009){Verbiest}, {Bailes}, {Coles}, {Hobbs}, {van
  Straten}, {Champion}, {Jenet}, {Manchester}, {Bhat}, {Sarkissian}, {Yardley},
  {Burke-Spolaor}, {Hotan}, \& {You}}]{vbc+09}
{Verbiest} J.~P.~W., {Bailes} M., {Coles} W.~A., {Hobbs} G.~B., {van Straten}
  W., {Champion} D.~J., {Jenet} F.~A., {Manchester} R.~N., {Bhat} N.~D.~R.,
  {Sarkissian} J.~M., {Yardley} D., {Burke-Spolaor} S., {Hotan} A.~W., {You}
  X.~P., 2009, 400, 951

\bibitem[{{Verbiest} {et~al.}(2008){Verbiest}, {Bailes}, {van Straten},
  {Hobbs}, {Edwards}, {Manchester}, {Bhat}, {Sarkissian}, {Jacoby}, \&
  {Kulkarni}}]{vbv+08}
{Verbiest} J.~P.~W., {Bailes} M., {van Straten} W., {Hobbs} G.~B., {Edwards}
  R.~T., {Manchester} R.~N., {Bhat} N.~D.~R., {Sarkissian} J.~M., {Jacoby}
  B.~A., {Kulkarni} S.~R., 2008, 679, 675

\bibitem[{Verbiest {et~al.}(2010)Verbiest, Bhat, Burke-Spolaor, Champion,
  Coles, Hobbs, Hotan, Jenet, Khoo, Lee, Lommen, Manchester, Oslowsk, Reynolds,
  Sarkissian, van Straten, Yardley, \& You}]{vbb+10}
Verbiest J.~P.~W.~Bailes M., Bhat N.~D.~R., Burke-Spolaor S., Champion D.~J.,
  Coles W., Hobbs G.~B., Hotan A.~W., Jenet F., Khoo J., Lee K.~J., Lommen A.,
  Manchester R.~N., Oslowsk S., Reynolds J., Sarkissian J., van Straten W.,
  Yardley D.~R.~B., You X.~P., 2010, Classical and Quantum Gravity, 27, 084015

\bibitem[{Walker {et~al.}(2008)Walker, Koopmans, Stinebring, \& {van
  Straten}}]{wksv08}
Walker M.~A., Koopmans L.~V.~E., Stinebring D.~R., {van Straten} W., 2008, 388,
  1214

\bibitem[{Williamson(1973)}]{wil73}
Williamson I.~P., 1973, 163, 345

\bibitem[{You {et~al.}(2007)You, Hobbs, Coles, Manchester, Edwards, Bailes,
  Sarkissian, Verbiest, {van Straten}, Hotan, Ord, Jenet, Bhat, \&
  Teoh}]{yhc+07}
You X.-P., Hobbs G., Coles W., Manchester R.~N., Edwards R., Bailes M.,
  Sarkissian J., Verbiest J.~P.~W., {van Straten} W., Hotan A., Ord S., Jenet
  F., Bhat N.~D.~R., Teoh A., 2007, 378, 493

\end{thebibliography}

\end{document}